\documentclass[pra,twocolumn,superscriptaddress]{revtex4-2}

\usepackage{nicefrac,amsmath,amssymb,mathrsfs}
\usepackage{graphicx}
\usepackage{natbib}
\usepackage{bm,color}
\usepackage{xspace}
\usepackage{tabularx}
\usepackage{fancyvrb}

\newcommand{\bem}[0]{\textsc{bem}\xspace}
\newcommand{\nanobem}[0]{\textsc{nanobem}\xspace}
\newcommand{\iscat}[0]{i\textsc{scat}\xspace}
\newcommand{\cobri}[0]{\textsc{cobri}\xspace}

\DefineVerbatimEnvironment{code}{Verbatim}{fontsize=\footnotesize,baselinestretch=1}
\DefineVerbatimEnvironment{boxcode}{Verbatim}{%
fontsize=\footnotesize,baselinestretch=1,frame=single,numbers=left}

\begin{document}

\title{Unified simulation platform for interference microscopy} 

\author{Felix Hitzelhammer}
\affiliation{Institute of Physics, University of Graz, Universit\"atsplatz 5, 8010 Graz, Austria}
\affiliation{University of Vienna, Faculty of Physics, VCQ, 1090 Vienna, Austria}
\affiliation{University of Vienna, Max Perutz Laboratories, Department of Structural and Computational Biology,
1030 Vienna, Austria}

\author{Anežka Dostálová}
\affiliation{Department of Optics, Faculty of Science, Palacký University, 17. listopadu 12, 77900 Olomouc, Czech Republic}
\affiliation{University of Vienna, Faculty of Physics, VCQ, 1090 Vienna, Austria}
\affiliation{University of Vienna, Max Perutz Laboratories, Department of Structural and Computational Biology,
1030 Vienna, Austria}

\author{Ilia Zykov}
\affiliation{University of Vienna, Faculty of Physics, VCQ, 1090 Vienna, Austria}
\affiliation{University of Vienna, Max Perutz Laboratories, Department of Structural and Computational Biology,
1030 Vienna, Austria}

\author{Barbara Platzer}
\affiliation{University of Vienna, Faculty of Physics, VCQ, 1090 Vienna, Austria}
\affiliation{University of Vienna, Max Perutz Laboratories, Department of Structural and Computational Biology,
1030 Vienna, Austria}

\author{Clara Conrad-Billroth}
\affiliation{University of Vienna, Faculty of Physics, VCQ, 1090 Vienna, Austria}
\affiliation{University of Vienna, Max Perutz Laboratories, Department of Structural and Computational Biology,
1030 Vienna, Austria}

\author{Thomas Juffmann}
\affiliation{University of Vienna, Faculty of Physics, VCQ, 1090 Vienna, Austria}
\affiliation{University of Vienna, Max Perutz Laboratories, Department of Structural and Computational Biology,
1030 Vienna, Austria}

\author{Ulrich Hohenester}
\affiliation{Institute of Physics, University of Graz, Universit\"atsplatz 5, 8010 Graz, Austria}
\thanks{Corresponding authors.  E-mail \texttt{ulrich.hohenester@uni-graz.at}.}

\date{\today}

\begin{abstract}
Interferometric scattering microscopy is a powerful technique that enables various applications, such as mass photometry and particle tracking. Here we present a numerical toolbox to simulate images obtained in interferometric scattering, coherent bright-field, and dark-field microscopy. The scattered fields are calculated using a boundary element method, facilitating the simulation of arbitrary sample geometries and substrate layer structures. A fully vectorial model is used for simulating the imaging setup. We demonstrate excellent agreement between our simulations and experiments for different shapes of scatterers and excitation angles. Notably, for angles near the Brewster angle, we observe a contrast enhancement which may be beneficial for nanosensing applications. The software is available as a \textsc{matlab} toolbox. 
\end{abstract}
\maketitle

\section{Introduction}

Interferometric scattering microscopy (\iscat) allows for the detection of nanoparticles with unprecedented sensitivity at high spatio-temporal resolution by exploiting the interference between scattered light from a sample and a reference beam. The technique has rapdily developed in recent years, finding numerous applications in biosciences and medicine, including the characterization of single proteins~\cite{Young2018QuantitativeMacromolecules}, and the tracking of metal nanoparticles on microsecond timescales~\cite{Taylor2019InterferometricMembrane}. Related techniques, such as coherent brightfield imaging (\cobri)~\cite{huang:17a,huang:17b} and dark-field imaging can in principle yield similar sensitivities~\cite{Dong2021FundamentalPhotometry}.

After initial proof-of-concept experiments, one of the primary goals has been to push the sensitivity limits to detect smaller and smaller particles, eventually enabling high-precision mass photometry of proteins~\cite{Young2018QuantitativeMacromolecules}. For such applications, \iscat images can be accurately simulated using analytical models~\cite{mahmoodabadi:20,Dong2021FundamentalPhotometry}. These models have been shown to accurately predict \iscat contrast levels and provide insight into the effects of aberrations, the z-dependence of the expected contrast, and the shot-noise limited maximally achievable localization and mass estimation precision. However, they typically only consider a single point dipole, or a spherical scatterer, inside a homogeneous medium. 

Interferometric imaging techniques are often used for tracking particles in cellular environments~\cite{Taylor2019InterferometricMembrane} or for sensing on top of structured substrates~\cite{Li2021Photonicresonator}. Furthermore, it has been demonstrated that high-precision quantitative agreement between experiment and simulation can only be achieved when the roughness of cover slips~\cite{Lin2022OpticalFingerprint}, and plasmonic enhancements~\cite{Shi2020ResonantScattering} are taken into account. Such complex geometries are difficult to simulate analytically and therefore require numerical approaches~\cite{he:21,Shi2020ResonantScattering,mahmoodabadi:20}. 

Here, we simulate interferometric microscopy images using the boundary element method (\bem) \cite{chew:95,hohenester:20} and its implementation within the \nanobem toolbox~\cite{hohenester.cpc:22,nanobem23}. Details on the software development are provided in the Supporting Information.  Within numeric constraints, arbitrary scattering geometries can be implemented concerning nanoparticle shape, nanoparticle distribution, substrate layer structure, illumination direction, and illumination polarization. The scattered far-fields are calculated and propagated to the camera using a fully vectorial imaging model based on the Richards-Wolf approach~\cite{richards:59}. This enables the simulation of high numeric aperture (NA) setups, including the most common \iscat geometries with both full-field or scanning excitation as well as the transmissive full-field approach, which we call \cobri in the following. We simulate \iscat images for off-axis illumination and demonstrate a contrast enhancement and tunability close to the Brewster angle. In addition, we find excellent agreement between our simulated results and experimental \iscat data for a gold nanosphere and silver nanocube.  


First, in Sec.~\ref{sec:theory}, we present the theory underlying interference microscopy, the \bem approach, and the Richards-Wolf approach. Selected simulation results and a comparison with experiments are given in Sec.~\ref{sec:results}. Finally, we provide a brief summary in Sec.~\ref{sec:summary}.  The simulation software and a short user manual can be found in the Supporting Information.

\section{Theory}\label{sec:theory}

\subsection{Outline of the problem}

\begin{figure}
\centerline{\includegraphics[width=\columnwidth]{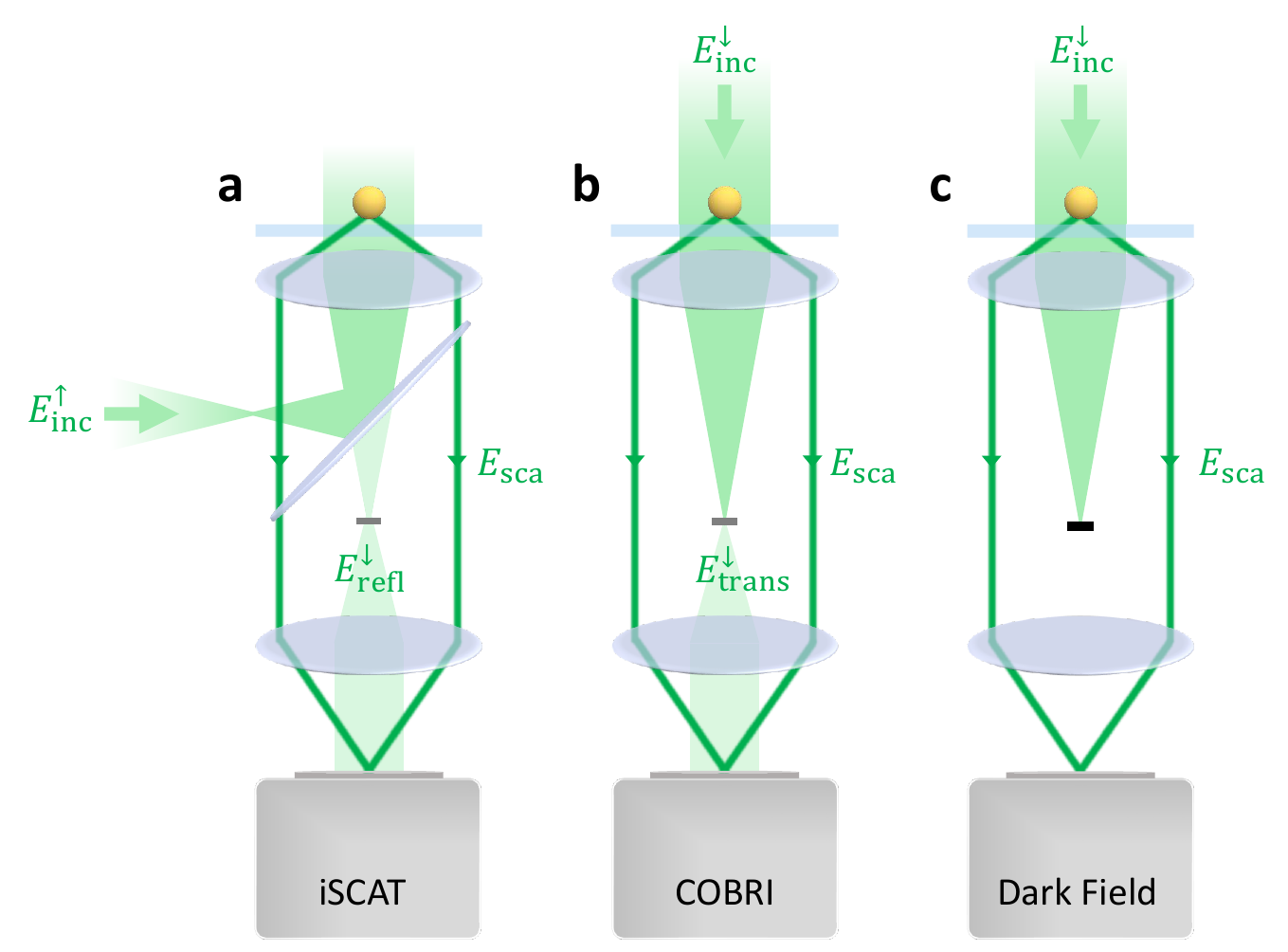}}
\caption{Sketch of microscopy configurations. (a) In \iscat, the interference image is formed between the (possibly attenuated) field reflected at the coverslip and the fields scattered by the nanoparticle.  (b) In \cobri, the interference image between the (possibly attenuated) transmitted incoming and scattered fields is measured. (c) In dark field microscopy, only the scattered fields are imaged. Figure adapted from~\cite{Dong2021FundamentalPhotometry}.}
\label{fig:sketch}
\end{figure}

We consider the situation depicted in Fig.~\ref{fig:sketch} where a nanoparticle located above a substrate is excited by a monochromatic incoming laser, and the interference between the (reflected or transmitted) incoming light and the light scattered by the nanoparticle is imaged through a microscopy setup.  In \iscat, the nanoparticle is excited from below, and the total incoming field is the sum of the primary field $\bm E_{\rm inc}^\uparrow$ impinging from below on the substrate, and the secondary fields resulting from reflection and transmission at the interface,
\begin{equation}\label{eq:einciscat}
  \mbox{\iscat:}\qquad \bm E_{\rm inc}=
  \begin{cases}
    \phantom{\bm E_{\rm inc}^\uparrow+}\,\,\bm E_{\rm trans}^\uparrow & \mbox{for $z>0$} \\ 
    \bm E_{\rm inc}^\uparrow+\bm E_{\rm refl}^\downarrow & \mbox{for $z<0$.}
  \end{cases}
\end{equation}
The superscript arrows indicate the propagation direction of the fields, given with respect to the $z$ direction.  In \cobri, the primary field $\bm E_{\rm inc}^\downarrow$ impinges on the interface from above, and the total incoming field is again the sum of primary and secondary fields
\begin{equation}\label{eq:einccobri}
  \mbox{\cobri:}\qquad \bm E_{\rm inc}=
  \begin{cases}
    \bm E_{\rm inc}^\downarrow+\bm E_{\rm refl}^\uparrow & \mbox{for $z>0$} \\ 
    \phantom{\bm E_{\rm inc}^\downarrow+}\,\,\bm E_{\rm trans}^\downarrow & \mbox{for $z<0$.}
  \end{cases}
\end{equation}
Eqs.~(\ref{eq:einciscat},\ref{eq:einccobri}) can be generalized for stratified media, i.e., vertically stacked materials with multiple interfaces, supplementing the incoming fields by the proper fields inside the various media.  See~\cite{chew:95,hohenester:20} for a detailed discussion of how to compute these fields.  In the following, we will assume that $\bm E_{\rm inc}$ is known.

In the presence of a nanoparticle, the incoming fields become scattered by the nanoparticle, and the total field is the sum of incoming and scattered fields
\begin{equation}
  \bm E_{\rm tot}=\bm E_{\rm inc}+\bm E_{\rm sca}\,.
\end{equation}
Far away from the nanoparticle and the interfaces of the stratified medium, the fields pass through the lenses of an imaging system, as will be described in more detail below and are transformed according to 
\begin{equation*}
  \mbox{imaging:}\qquad\bm E_{\rm inc}+\bm E_{\rm sca}\longrightarrow\bm E_{\rm inc}'+\bm E_{\rm sca}'\,.
\end{equation*}
The image on the camera is then proportional to the field intensity, and we get
\begin{eqnarray}\label{eq:image}
  \mbox{\iscat:} &&\qquad \mathcal{I}_{\rm iSCAT}\propto
  \left|\bm E_{\rm refl}^{\downarrow\prime}+\bm E_{\rm sca}^{\downarrow\prime}\right|^2 \nonumber\\
  \mbox{\cobri:} &&\qquad \mathcal{I}_{\rm COBRI}\propto
  \left|\bm E_{\rm trans}^{\downarrow\prime}+\bm E_{\rm sca}^{\downarrow\prime}\right|^2 \,.
\end{eqnarray}

When describing \iscat or \cobri theoretically, usually the different image formation steps are modeled in consecutive order and with various degrees of sophistication.  For incoming plane waves and a substrate, the reflected and transmitted waves can be obtained with the usual Fresnel coefficients~\cite{novotny:06}.  The nanoparticle is usually modeled as a polarizable particle with homogeneous material parameters, and the light scattered by the nanoparticle is approximated by the emission of a point dipole~\cite{mahmoodabadi:20}.  Finally, the imaging of the secondary and scattered fields is described through the Richards and Wolf approach~\cite{richards:59}, which we will discuss in more detail below.  While for small nanoparticles, the agreement between the results of such an approach and experiment are usually very good~\cite{mahmoodabadi:20,Dong2021FundamentalPhotometry}, deviations might occur for larger, coated or coupled nanoparticles, as well as for tilted or structured incoming light fields.  In such cases, a general Maxwell solver should be used that allows computing the scattered electromagnetic fields for arbitrary incoming fields and nanoparticle geometries.

\begin{table*}
\caption{Field manipulations needed to render the \nanobem toolbox suitable for the simulation of interference microscopy.  For the representation of the electromagnetic fields, we consider far-fields and plane-wave decompositions.  For details see text.}\label{tab:iscat}
\begin{tabularx}{2\columnwidth}{llX}
\hline\hline
Object & Transformation & Description \\
\hline\\[-8pt]
Rotation matrix $R$ & $R{\bm F}$, $R\hat{\bm r}$ & Rotate far-field amplitude and propagation direction \\
& $R{\bm\epsilon}$, $R{\bm k}$ & Rotate field and wavevector for plane-wave decomposition \\
Shift vector $\bm r_0$ & $e^{-ik\hat{\bm r}\cdot\bm r_0}\bm F$ & Shift coordinate system of far-field expansion  from $\bm r\to\bm r+\bm r_0$ \\
& $e^{-i\bm k\cdot\bm r_0}\bm\epsilon$ & Shift coordinate system of plane-wave decomposition \\
Imaging lenses & $\frac{ik}{2\pi}\mathcal{I}_{\rm im}\left[\bm F(\hat{\bm r})\right]$ & Image far-fields using Richards-Wolf approach\\
& $\phantom{\frac{ik}{2\pi}}\,\,\mathcal{I}_{\rm im}\big[\bm \epsilon(\hat{\bm k})\big]$ & Image plane-wave decomposition using Richards-Wolf approach\\[2pt]
\hline
\hline
\end{tabularx}
\end{table*}

\subsection{Ingredients of interference microscopy toolbox}

In this work, we employ the open-source toolbox \nanobem that is based on a boundary element method (\bem) approach and is implemented in \textsc{matlab}~\cite{hohenester.cpc:22,nanobem23}.  To render the toolbox suitable for the simulation of various interference microscopy techniques, additional features are required, which are listed in Table~\ref{tab:iscat}.  We consider a single plane wave or a plane-wave decomposition of focused laser fields for the incoming fields and a far-field representation for the scattered fields.  For both types of field representation, we have implemented imaging transformations based on the Richards-Wolf approach~\cite{richards:59}.  To allow for the rotation of the optical axis and the shift of the focus position, we require additional operations that enable the rotation and shifting of the field representations.  These operations are usually encapsulated in the toolbox functions and are discussed at some length in the Supporting Information.

For the incoming fields, we use a plane-wave decomposition 
\begin{equation}\label{eq:decompose}
  \bm E_{\rm inc}(\bm r)=\int e^{i\bm k\cdot\bm r}\bm\epsilon_{\rm inc}(\hat{\bm k})\,d{\rm\Omega}\,,
\end{equation} 
where $\bm\epsilon_{\rm inc}(\hat{\bm k})$ is the field amplitude for the propagation direction $\hat{\bm k}$, and $d{\rm\Omega}$ denotes the integration over the unit sphere.  With this, we can also describe a plane wave with polarization $\bm\epsilon_0$ and propagation direction $\hat{\bm k}_0$ by setting $\bm\epsilon_{\rm inc}(\hat{\bm k})=\bm\epsilon_0\,\delta(\cos\theta-\cos\theta_0)\delta(\phi-\phi_0)$, where $\theta_0$, $\phi_0$ are the polar and azimuthal angles of the light propagation direction, respectively.  For the scattered fields, we use a far-field representation in terms of outgoing spherical waves modulated by the far-field amplitude $\bm F_{\rm sca}(\hat{\bm r})$~\cite[Eq.~(3.5)]{hohenester:20},
\begin{equation}\label{eq:farfield}
  \bm E_{\rm sca}(\bm r)\xrightarrow[kr\gg 1]{} \frac{e^{ikr}}r \bm F_{\rm sca}(\hat{\bm r})\,.
\end{equation}
In our simulation approach, primary incoming fields are specified, which are typically either plane waves $\bm\epsilon_{\rm inc}^{(0)}(\hat{\bm k}_0)$ or the focal fields of an incoming laser field~\cite[Eq.~(3.47)]{novotny:06}

\begin{displaymath}
  \bm E_{\rm laser}\xrightarrow[\rm Focus]{}\bm\epsilon_{\rm inc}^{(0)}(\hat{\bm k})\,.
\end{displaymath}
See the Supporting Information for more details of how to obtain these fields.  In a second step, the fields are used as an input for the \bem solver, which computes the secondary incoming fields $\bm\epsilon_{\rm inc}^{(1)}$ and the scattered far-fields
\begin{displaymath}
  \bm\epsilon_{\rm inc}^{(0)}(\hat{\bm k})\longrightarrow\mbox{\bem solver}\longrightarrow
  \bm\epsilon_{\rm inc}^{(1)}(\hat{\bm k})\,,\,\bm F_{\rm sca}(\hat{\bm r})\,,
\end{displaymath}
which are finally submitted to the imaging procedure of the Richards-Wolf approach. We then obtain the \iscat or \cobri images through
\begin{displaymath}
  \bm\epsilon_{\rm inc}^{(1)}(\hat{\bm k})\,,\,\bm F_{\rm sca}(\hat{\bm r})\longrightarrow\mbox{Imaging}\longrightarrow
  \left|\bm E_{\rm inc}^{(1)'}+\bm E'_{\rm sca}\right|^2\,.
\end{displaymath}

\begin{figure*}
\centerline{\includegraphics[width=1.5\columnwidth]{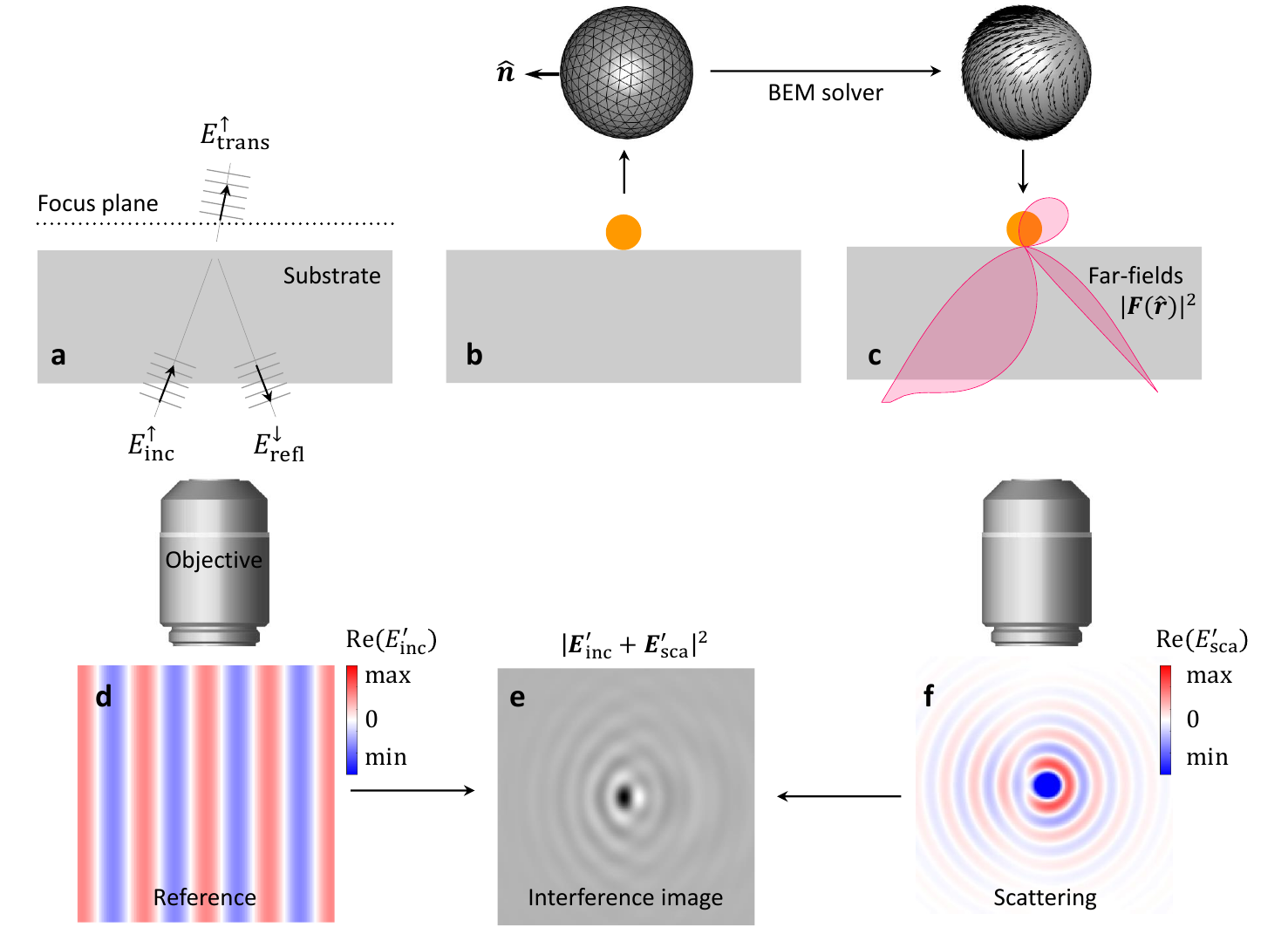}}
\caption{(a) In our simulations, we consider a substrate or stratified medium that is excited by incoming fields.  (b) The optical response of an additional nanoparticle, here a gold nanosphere, is obtained from a \bem solver.  The boundary of the nanoparticle is discretized through triangular boundary elements, and by solving the \bem working equations, the tangential electromagnetic fields at the particle boundary, which fully characterize the solution, are obtained.  (c) From the \bem solutions we can compute the optical far-fields, which are imaged using the Richards-Wolf approach.  Through the superposition of the imaged (d) reference and (f) scattering fields we finally obtain the (e) interference image that can be directly compared with results of \iscat experiments.}
\label{fig:bemschematics}
\end{figure*}

\subsection{BEM in a nutshell}\label{sec:bemshort}

In \bem nanoparticles with homogeneous material parameters, which are separated by abrupt interfaces from the embedding medium, are considered.  The shape of the particles can be chosen arbitrarily, and coupled or coated particles modeled accurately.  Whether the restriction of homogeneous media applies to a given system under study must be decided on a case-by-case basis.  In the following, we briefly introduce the methodology of \bem, more details can be found in the specialized literature~\cite{chew:95,hohenester:20}.

Consider a nanoparticle embedded in a background medium with the permittivity and permeability functions $\varepsilon$, $\mu$, respectively.  The Stratton-Chu~\cite{stratton:39} or representation formula~\cite[Eq.~(5.37)]{hohenester:20} then relates the scattered electric field at position $\bm r$ inside the background medium to the tangential electromagnetic fields at the particle boundary via
\begin{equation}\label{eq:representation}
  \bm E_{\rm sca}(\bm r)=\oint\Big(i\omega\mu\,{\rm SL}[\hat{\bm n}'\times\bm H'](\bm r)-
  {\rm DL}[\hat{\bm n}'\times\bm E'](\bm r)\Big)\,dS'\,.
\end{equation}
The integration extends over the nanoparticle boundary and $\hat{\bm n}$ is the surface normal pointing away from the particle.  The single and double layer potentials $\rm SL$, $\rm DL$ are related to the dyadic Green's tensor for the stratified medium and describe how the tangential fields propagate from the nanoparticle boundary at position $\bm s'$ to $\bm r$.

The representation formula of Eq.~\eqref{eq:representation} can be used in two different ways.  First, by letting $\bm r\!\to\!\bm s$ approach a position on the boundary, we obtain the so-called Calderon identity, which relates the electric field $\bm E(\bm s)$ on the boundary (left-hand side) to an integral over the tangential boundary fields (right-hand side).  Combining the Calderon identities for the electromagnetic fields outside and inside the particle with the usual Maxwell boundary conditions of continuous tangential electromagnetic fields, we obtain an expression that allows computing the tangential boundary fields for a given incoming field.  This expression forms the heart of computational \bem solvers~\cite[Eq.~(11.41)]{hohenester:20}.  Second, knowing the tangential fields on the boundary, we can use Eq.~\eqref{eq:representation} to compute the fields away from the boundary.  Because the layer potentials $\rm SL$, $\rm DL$ describe the field propagation in the stratified medium, we can perform the far-field limit analytically to arrive at
\begin{equation}
  {\rm SL}[\hat{\bm n}\times\bm H](\bm r)\xrightarrow[kr\gg 1]{} \left(\frac{e^{ikr}}r\right)
  \mathcal{S}[\hat{\bm n}\times\bm H](\hat{\bm r})\,,
\end{equation}
with a similar expression for the double-layer potential~\cite{hohenester:20}.  Here $\mathcal{S}$ is a function that depends on the tangential boundary fields and the propagation directions $\hat{\bm r}$ only.  

Fig.~\ref{fig:bemschematics} shows the schematics of a typical \bem simulation.  (a) We start by defining a reference structure, here a substrate, and the incoming fields, here a plane wave impinging from below on the interface.  Because of the contrast in refractive indices, part of the wave is reflected and transmitted at the interface.  (b) The modification of the electromagnetic fields in the presence of a nanoparticle, here a gold nanosphere, is accounted for through the \bem solver: the boundary of the nanoparticle is discretized in terms of triangular boundary elements, and by solving the \bem working equation~\cite[Eq.~(9)]{nanobem23} the tangential boundary fields that fully characterize the solution of the problem under study are obtained.  (c) By submitting the far-fields to an imaging transformation, as described in more detail in Sec.~\ref{sec:lens}, we obtain the fields in the image plane of the objective.  The superposition of the (f) scattered and (d) reference (reflected) fields allows us to finally compute the (e) interference image, see Eq.~\eqref{eq:image}, which can be directly compared with the results of \iscat or \cobri experiments.

\subsection{Richards-Wolf approach}\label{sec:lens}

\begin{figure*}[t]
\includegraphics[width=1.85\columnwidth]{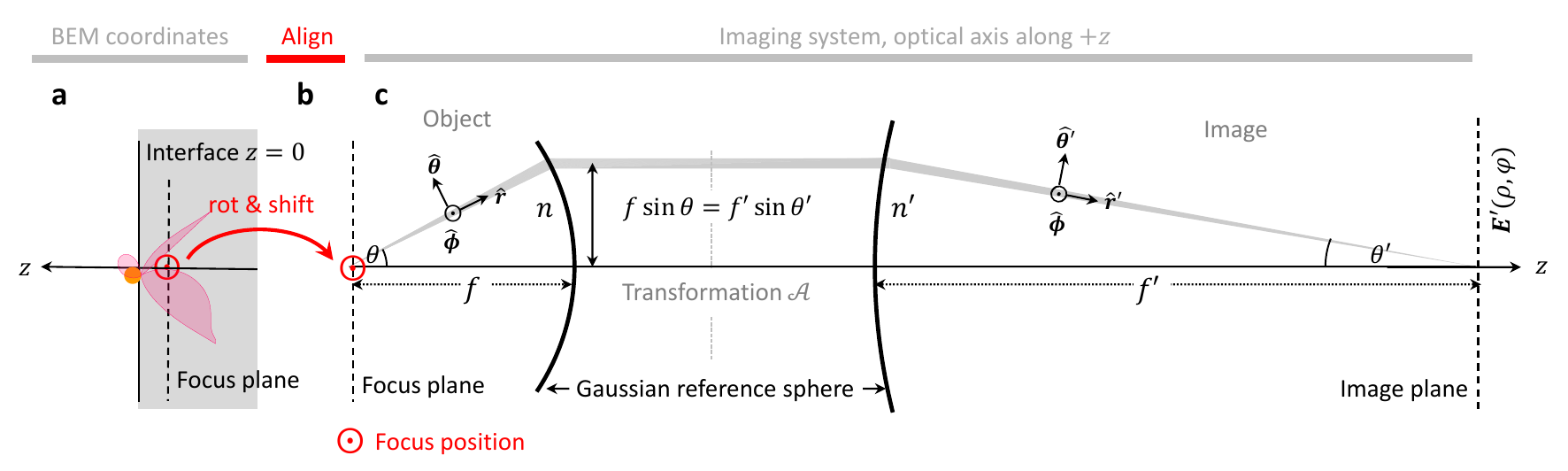}
\caption{Imaging through an aplanatic lens.  (a) \bem simulation using a coordinate system where $z=0$ coincides with the interface of the substrate. (b) Through a rotation and shift of the fields, we transform them to the (c) coordinate system for imaging, with the focus point at $z=0$ and the optical axis pointing along $+z$.  The far-fields emanating from the focus position are imaged in the image plane.  In the Richards-Wolf approach, the light rays are redirected at the Gaussian reference spheres, and we additionally consider the conservation of energy transported by the waves.  We use unprimed spherical coordinates on the object side and primed coordinates on the image side.  Details can be found in the text.  A transformation $\mathcal{A}$ in the back focal plane between the two reference spheres allows us to manipulate the fields and mimic, e.g., the effect of a quarter-wave plate.}
\label{fig:lens}
\end{figure*}

The working principle of the Richards-Wolf approach for imaging has been described at length elsewhere~\cite{richards:59,novotny:06,khadir:19,hohenester:20,mahmoodabadi:20}. Here, we only sketch the main steps. The approach requires the secondary incoming fields and the scattered far-fields as input and gives the imaged fields as output.  The imaging of the objective is modeled by two Gaussian reference spheres with radii $f$ and $f'$, as shown in Fig.~\ref{fig:lens}(c).  On the object side, the fields emanate from the focus point and become redirected at the Gaussian reference spheres.  As a result of these operations, we obtain an image of the fields in the focus plane that is magnified by a factor of
\begin{equation}\label{eq:magnification}
  M=\frac n{n'}\frac{f'}f\,,
\end{equation}
where the refractive indices on the object and image side are indicated with unprimed and primed symbols $n$, $n'$, respectively.  The same convention of unprimed (object) and primed (image) symbols is adopted for all other quantities, such as coordinates or vectors.

Let us pause for a moment and clarify a point that can cause some confusion, namely the location of the focus spot.  Suppose first that in Fig.~\ref{fig:lens}(c), the space left to the first reference sphere (with focus $f$) is filled with a homogeneous medium with refractive index $n$, for which the microscope objective was designed, e.g., oil, and $\bm r_{\rm foc}$ is the focus spot that forms the origin of the coordinate system of the imaging system.  If we send a light beam propagating parallel to the optical axis in the opposite direction back to the first reference sphere, it will be focused at $\bm r_{\rm foc}$.  When replacing the homogeneous medium with a stratified medium, where the medium closest to the reference sphere has the refractive index $n$ again, we keep $\bm r_{\rm foc}$ fixed.  However, a light beam propagating back through the reference sphere does not necessarily focus at $\bm r_{\rm foc}$ but might be diffracted because of the stratified medium.  In this respect, $\bm r_{\rm foc}$ a hypothetical focus point is needed in our theoretical approach but might be a point of no particular significance in the experiment.  As shown in Fig.~\ref{fig:lens}(a), the coordinate systems of the \bem simulation and the imaging system can also differ.  In \bem, $z=0$ is usually defined as the upper interface position of the stratified medium. In contrast, for the imaging system, the center is located at the focus point, and the optical axis is assumed to point in the $+z$-direction.  To transform from one coordinate system to the other, we have to rotate and shift the fields, as discussed in more detail in the Supporting Information. Shifting displaces the focus point with respect to the \bem coordinate system, corresponding to the defocusing procedure conveniently employed in \iscat experiments.

In the Richards-Wolf approach, we trace the fields through the optical setup and ensure energy conservation.  For simplicity, we consider lenses with antireflecting coatings, thus neglecting back-reflections, and perfect imaging properties.  In principle, both approximations can be easily lifted, see for instance~\cite[Sec.~3.6]{hohenester:20} or \cite[Eq.~(14)]{mahmoodabadi:20}.  On the object side, we introduce a spherical coordinate system with the unit vectors $\hat{\bm\theta}$, $\hat{\bm\phi}$, $\hat{\bm r}$. Since the far-fields propagate in the direction $\hat{\bm r}$, the transverse electromagnetic fields only have components in the $\theta$, $\phi$ directions.  The $\phi$ component of the electric field has a \textsc{te} character and thus keeps its orientation when propagating through the optical system; see also Fig.~\ref{fig:lens}(c).  In contrast, the $\theta$ component has a \textsc{tm} character, and is rotated from the $\hat{\bm\theta}$ to the $\hat{\bm\theta}'$ direction.  With the additional requirement of the conservation of energy transported by the waves~\cite[Eq.~(3.30)]{hohenester:20}, we are then led to a relation between the electric fields on the object and image side via
\begin{equation}
  \bm E'=\sqrt{\frac{n}{n'}}\sqrt{\frac{\cos\theta'}{\cos\theta}}\,
  \mathcal{R}\cdot T[\bm E]\,,
\end{equation}

\begin{figure*}[t]
\includegraphics[width=1.8\columnwidth]{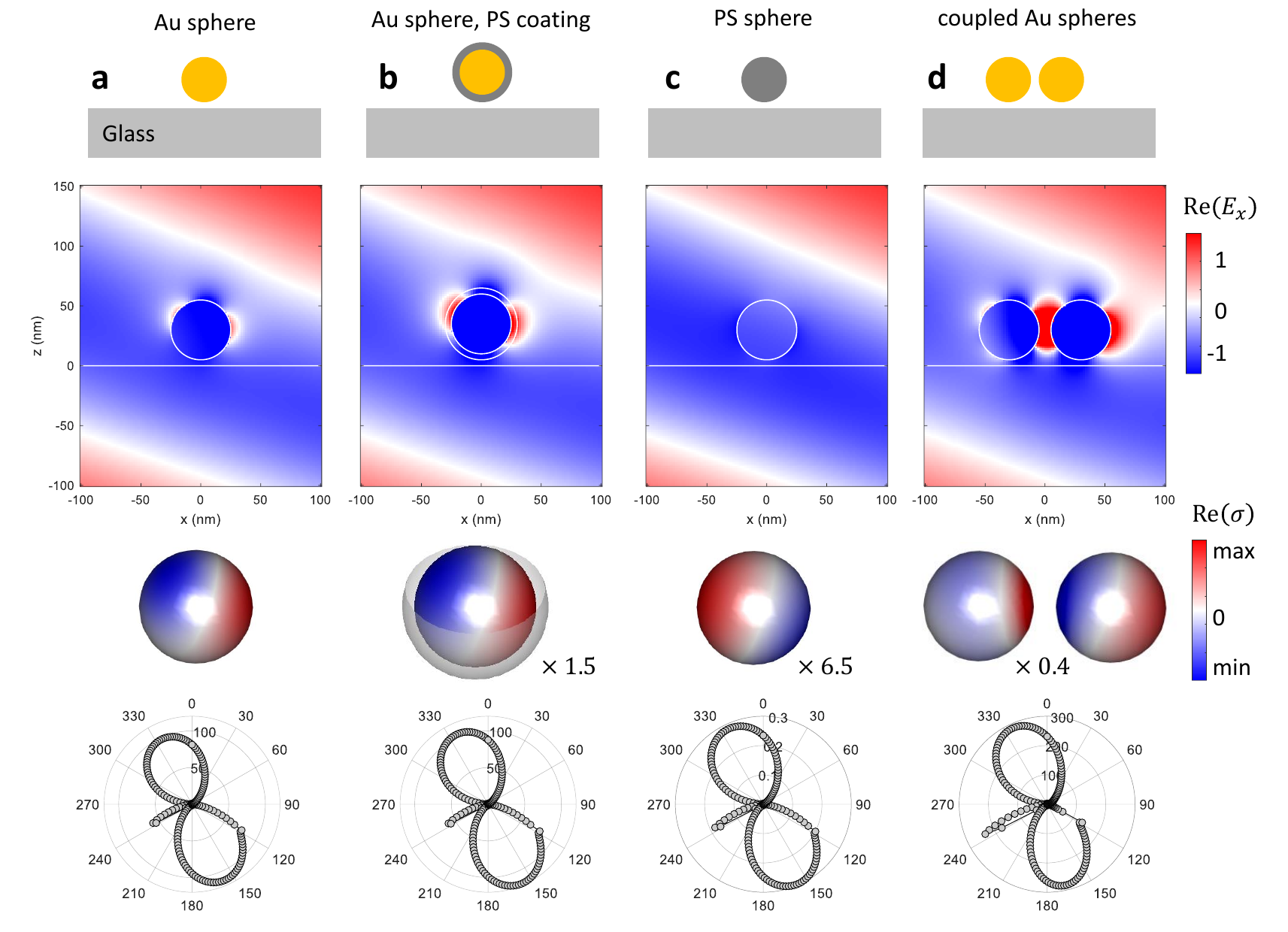}
\caption{Results of \bem simulations for (a) gold nanosphere (b), gold nanosphere coated with polystyrene (\textsc{ps}) (c) \textsc{ps} nanosphere, and (d) coupled gold nanospheres, embedded in water and situated 5 nm above a glass substrate. The sphere diameters are 50 nm, the coating thickness is 5 nm, and the gap distance is 10 nm. For the excitation, we use a plane wave with $\lambda=520$ nm and \textsc{tm} polarization that impinges on the interface from below with an angle of $-20^\circ$ with respect to the $z$-axis (anti-clockwise).  The second row shows the real part of the electric field map $E_x(x,z)$ in units of the incoming field, as obtained from the \nanobem toolbox (note that in panel (d) the field enhancement in the gap region is around six and exceeds the color limits), the third row shows the surface charge distributions (with scaling factors reported in the insets), and the last row shows the power scattered into a given far-field direction.}
\label{fig:bemsimulation}
\end{figure*}

\noindent with the rotation matrix $\mathcal{R}=\hat{\bm\phi}\hat{\bm\phi}+\hat{\bm\theta}'\hat{\bm\theta}$ that accounts for the rotation of the unit vectors from the object to the image side, and the transformation $T[\bm E]$ that accounts for the aforementioned change of coordinate systems.  In the last step, we assume $f'\gg f$ and perform the paraxial approximation by setting $\cos\theta'\approx 1$, such that $\hat{\bm r}'$ approximately points in the $z$-direction.  With this, we are led to the Richards-Wolf integral
\begin{widetext}
\begin{equation}\label{eq:richards-wolf}
  \mathcal{I}_{\rm im}\left[\bm\epsilon_{\rm inc}^{(1)}\right]=\sqrt{\frac n{n'}}\frac{e^{i\psi}}{M}
  \int_0^{\theta_{\rm max}}\sqrt{\cos\theta}\sin\theta\,d\theta
  \int_0^{2\pi}d\phi\,\mathcal{A}\Big(\mathcal{R}\cdot T\big[\bm\epsilon_{\rm inc}^{(1)}(\theta,\phi)\big]\Big)e^{ik\left(\frac\rho M\right)\sin\theta\cos(\phi-\varphi)}\,,
\end{equation}
\end{widetext}
where $\theta_{\rm max}$ is the opening angle of the first reference sphere, which is related to the numerical aperture of the lens via $n\sin\theta_{\rm max}={\rm NA}$, and $\psi$ is a phase that is of no importance for our present concern.  $\mathcal{A}$ is an optional function that can be used to transform the fields in the back focal plane, see Fig.~\ref{fig:lens}. It could, for example, be used to model for the effect of a quarter-wave plate~\cite{hecht:98}, an attenuator for contrast enhancement, or a beam block for dark-field microscopy.  For details, see the Supporting Information.  Expressing the image coordinates $x'$, $y'$ in polar coordinates $\rho$, $\varphi$, we get for the imaged secondary fields $\bm E_{\rm inc}'(\rho,\varphi)=\mathcal{I}_{\rm im}[\bm\epsilon_{\rm inc}(\hat{\bm k})]$.  Similarly, for the imaging of optical far-fields we get~\cite[Eq.~(3.10)]{hohenester:20}
\begin{equation}\label{eq:lensfar}
  \bm E_{\rm sca}'(\rho,\varphi)=\left(\frac{ik}{2\pi}\right)\mathcal{I}_{\rm im}\left[\bm F_{\rm sca}(\hat{\bm r})\right]\,.
\end{equation}
where $\bm F_{\rm sca}$ is the far-field amplitude on the object side.

\section{Results}\label{sec:results}

\begin{figure*}[t]
\includegraphics[width=1.6\columnwidth]{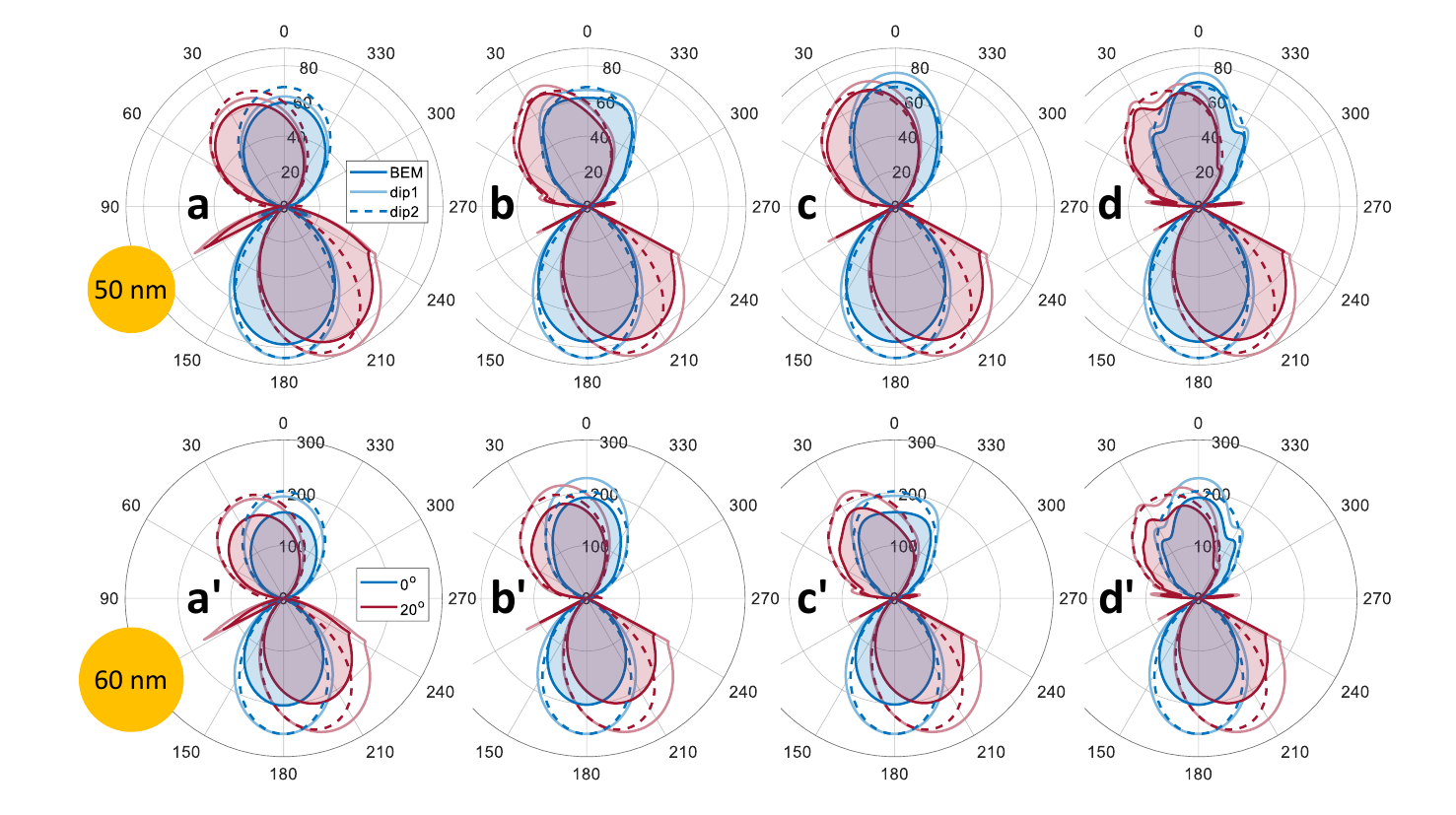}
\caption{Far-fields for gold spheres with different diameters and for gap distances between the sphere and glass-water interface of (a) 5 nm, (b) $\nicefrac\lambda 2$, (c) $\lambda$, and (d) $2\,\lambda$, with $\lambda=520$ nm.  In the simulations, a plane wave impinges on the interface from below with an angle of $0^\circ$ (blue lines) and $-20^\circ$ (red lines).  The polar plots indicate the power radiated in the far-field zone into a given direction.  We compare the results of the full \bem simulations (solid lines with shaded areas, \texttt{BEM}) with those obtained with the dipole approximation (faint solid lines, \texttt{dip1}), as described in more detail in the text, and a model where we consider dipole radiation in homogeneous space and consecutive diffraction of the emitted fields at the glass-water interface (dashed lines, \texttt{dip2}).}
\label{fig:farfields}
\end{figure*}

\subsection{BEM simulation}

We have implemented the interference microscopy features in our recently developed \nanobem toolbox~\cite{hohenester.cpc:22}, with the extension for stratified media using the matrix-friendly approach of Chew~\cite{nanobem23,chew:06,chew:09}.  See also the Supporting Information.  Although the details of the numerical implementation of the various toolbox functions are intricate, it is easy to set up and run simulations for simple geometries.  

Figure~\ref{fig:bemsimulation} shows simulation results for gold (Au) and polystyrene (\textsc{ps}) nanospheres.  Throughout, the spheres are embedded in water (refractive index $n=1.33$) and located 5 nm above a glass substrate ($n=1.5$).  The optical properties for Au are taken from~\cite{johnson:72}, and we use $n=1.59$ for \textsc{ps}.  We consider a laser excitation with a vacuum wavelength of 520 nm, where a plane wave with \textsc{tm} polarization impinges on the interface from below with an angle of $-20^\circ$ with respect to the $z$-axis (anti-clockwise).  In the second row of the figure, we show the electric field maps, which can be directly obtained with the \nanobem toolbox functions.  The computation of these maps is time-consuming because the fields have to be propagated from the sphere boundary to the field points $(x,z)$ by means of the representation formula of Eq.~\eqref{eq:representation}.  In contrast, the computation of the surface charges (third row) and the power emitted into a given far-field direction (last row) is significantly faster, because these quantities can be directly obtained from the tangential boundary fields alone.  Comparison of the results for metallic (Au) and dielectric (\textsc{ps}) nanospheres shows a phase difference of the surface charges.  This is because metallic nanoparticles sustain surface plasmon resonances in the optical regime~\cite{hohenester:20}, associated with coherent electron charge oscillations at the metal surface, which oscillate close to resonance with the usual $90^\circ$ phase delay with respect to the driving laser fields.  A large field enhancement in the gap region between the coupled spheres is observed, which is a well-known property of surface plasmon resonances~\cite{schuller:10}.  These results demonstrate that the \nanobem toolbox is well suited for the numerical solution of Maxwell's equations for a variety of different geometries and samples.  

In Fig.~\ref{fig:farfields} we show the emission patterns of gold nanospheres with different diameters and distances to the substrate interface.  We compare the results of the full \bem simulations (solid lines with shaded areas, \texttt{BEM}) with those for dipole approximations assuming a polarizable particle. In the latter, the nanoparticle scattering is either approximated by the scattering of a point dipole above an interface~\cite[Eq.~(D.6)]{novotny:06} (faint solid lines, \texttt{dip1}) or the scattering of a point dipole in homogeneous space followed by the diffraction of light at the substrate interface~\cite{mahmoodabadi:20} (dashed lines, \texttt{dip2}). We observe that the dipole approximation works well for nanoparticles diameters smaller than 50 nm.  For larger particles retardation effects gain importance and a full simulation approach becomes indispensable.  

\subsection{Simulation of imaging}

\begin{figure*}[t]
\includegraphics[width=1.7\columnwidth]{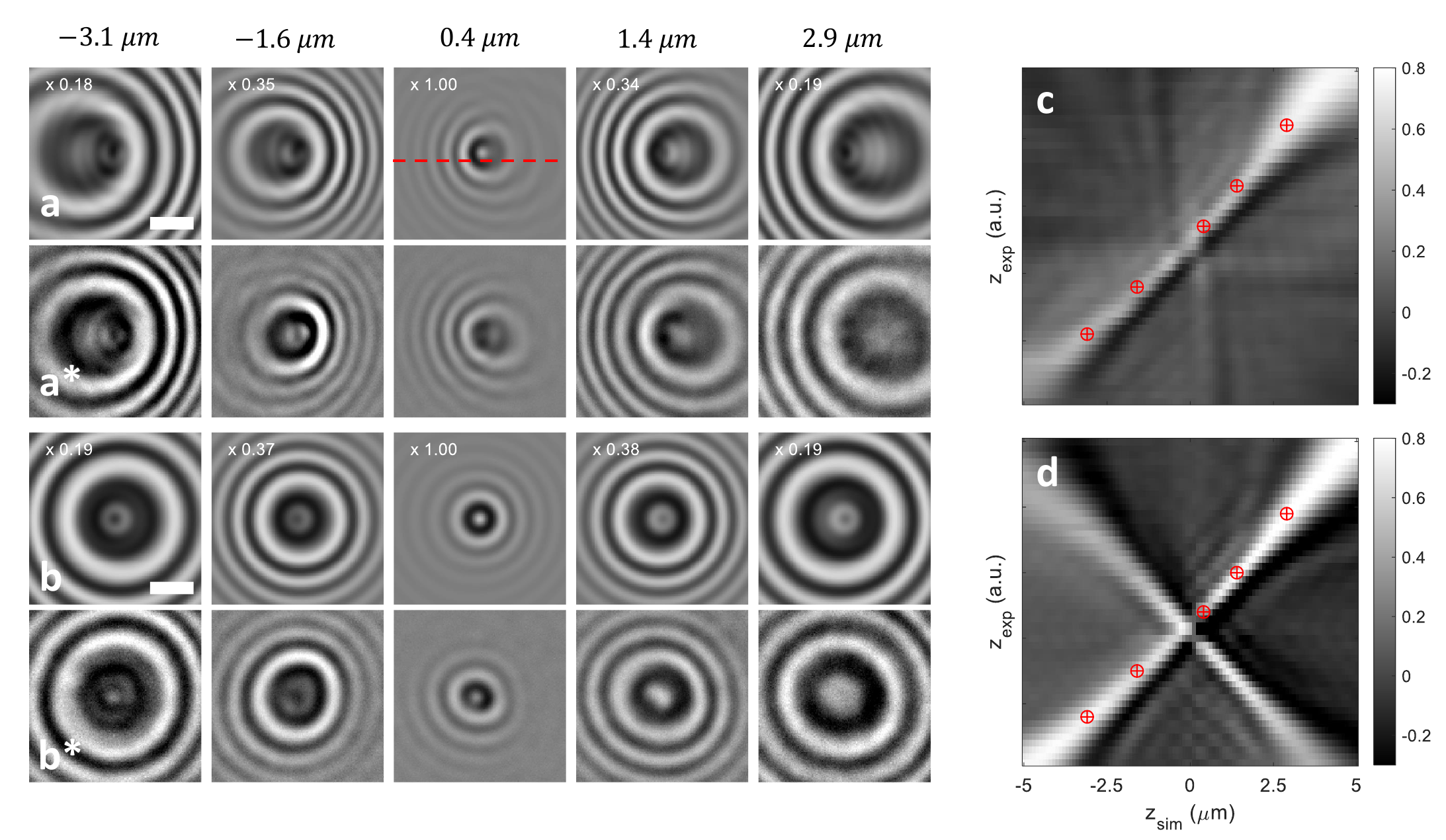}
\caption{(a,b) Simulated and (a*,b*) measured \iscat images for different focal plane positions reported above the panels.  We use the setup depicted in Fig.~\ref{fig:bemsimulation}(a), but for a gold nanosphere with a diameter of 55 nm, and for angles $\theta$ of the incoming plane wave with respect to the $+z$-axis of (a) $14^\circ$ and (b) $-2^\circ$.  The scale bar is $1 \, \mu$m, and the red dashed line indicates $y=0$.  (c,d) Correlation plots from which the best overlap between the simulated and measured images is obtained.  For details, see text.}
\label{fig:experiment}
\end{figure*}

In the previous section, we have discussed that the \nanobem toolbox allows computing the incoming fields and far-fields for typical interference microscopy setups without introducing any modifications.  Runtimes for typical setups range from a few tens of seconds to minutes on a normal desktop computer.  In what follows, we discuss the steps needed to compute the interference images, see Eq.~\eqref{eq:image}, that can be directly compared with the experiment.  In Appendix~\ref{sec:simulation}, we provide details about our numerical implementation of the Richards-Wolf approach.

We first discuss the simulation of the experimental setup depicted in Fig.~\ref{fig:setup1}, where the incoming laser beam passes through a quarter-wave plate (\textsc{wp}), and the focus lens, excites the nanoparticle, and the reflected and scattered fields are imaged.  The field manipulation of the quarter-wave plate is accounted for by a Jones matrix with the fast $x$-axis rotated clockwise by $45^\circ$~\cite{hecht:98}, as discussed in more detail in the Supporting Information, both in the excitation (light propagation in $+z$-direction) and detection (light propagation in $-z$-direction) path.  For imaging we apply the same Jones matrix in the $\mathcal{A}$ function, see Eq.~\eqref{eq:richards-wolf}, that allows for field manipulations in the back focal plane.

\begin{figure*}[t]
\includegraphics[width=2\columnwidth]{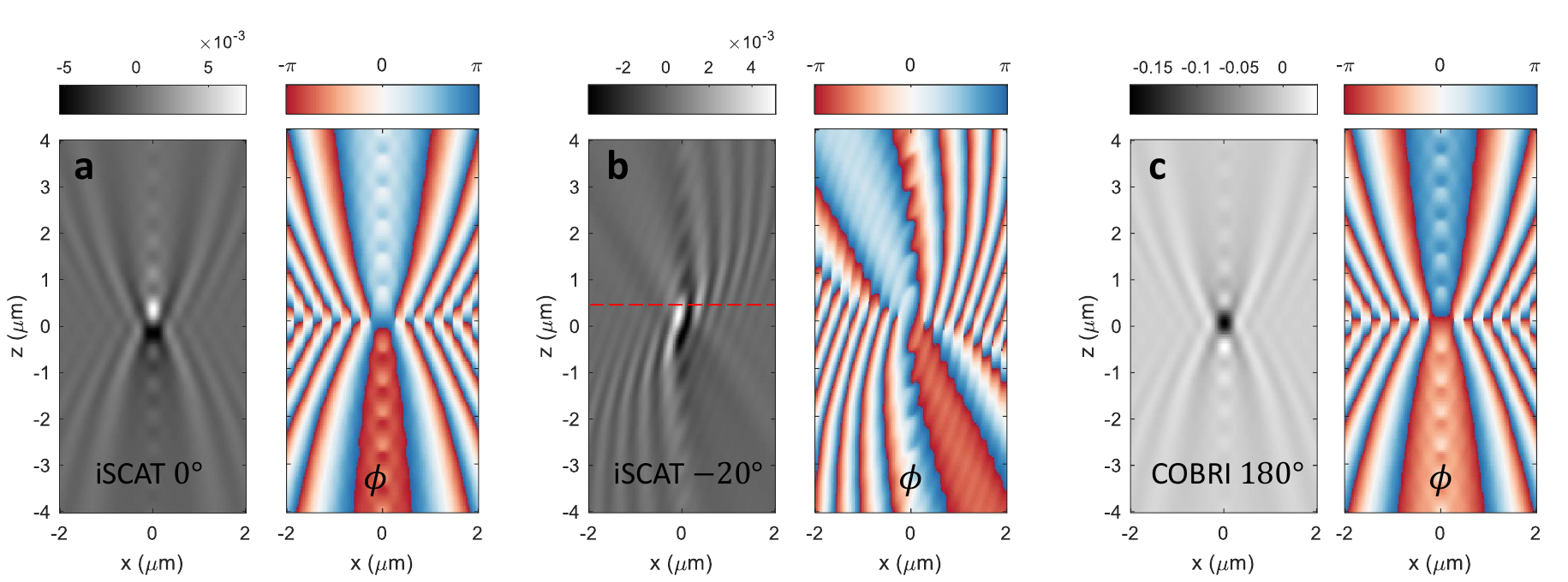}
\caption{\iscat $\mathcal{I}_{\rm sim}(x,y=0;z)$, phase $\phi=\arg (\bm E_{\rm ref}'\cdot\bm E_{\rm sca}^{\prime\,*})$, and \cobri maps for different focal positions $z$.  We use the same simulation setup as in Fig.~\ref{fig:bemsimulation}(a), and plane wave excitations with \textsc{tm} polarizations and incoming angles of (a) $0^\circ$, (b) $-20^\circ$, and (c) $180^\circ$ for \cobri.  Note that the plane of highest contrast, as indicated by the dashed red line, does not necessarily coincide with the focus plane of the objective.}
\label{fig:iPSF}
\end{figure*}

\begin{figure*}[t]
\centerline{\includegraphics[width=0.9\textwidth]{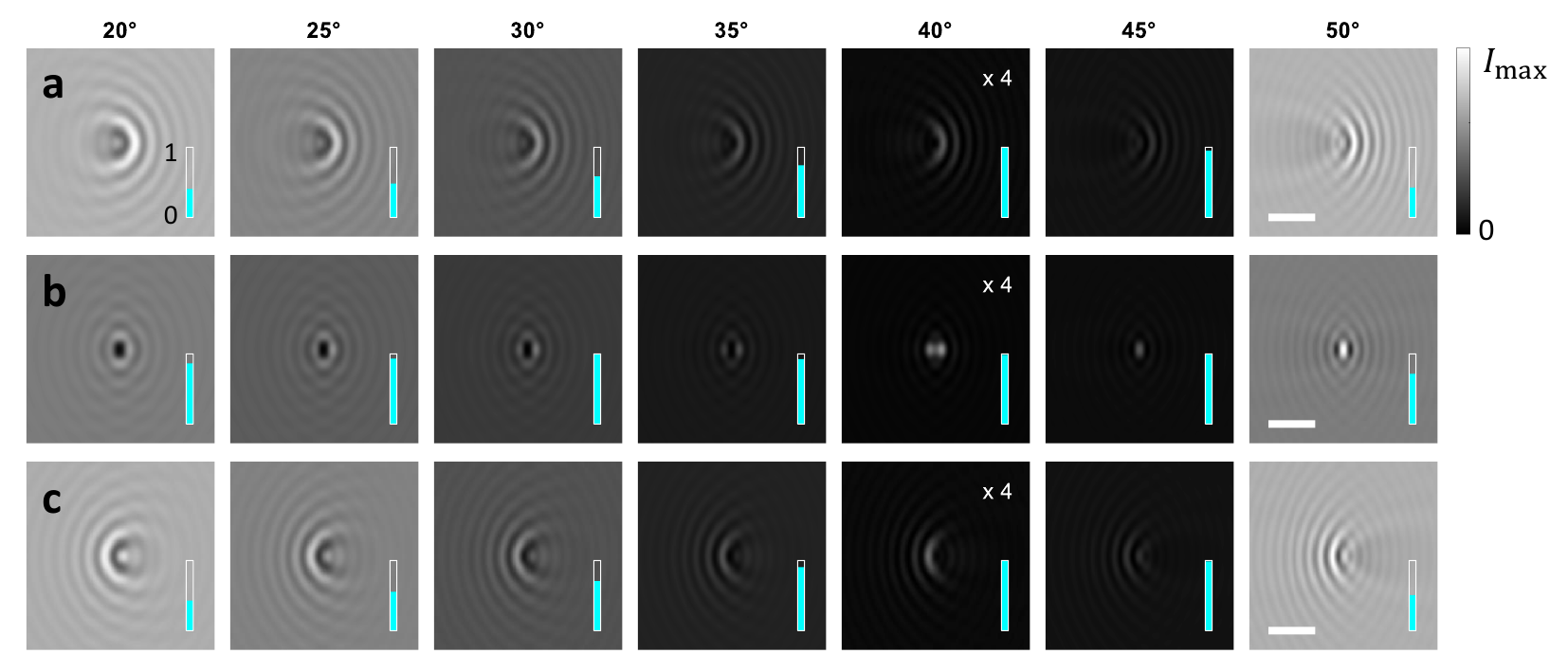}}
\caption{\iscat maps $\mathcal{I}^\theta(x,y;z_{\rm foc})$ without background subtraction for different incoming angles (see top) around the Brewster angle $\theta_B\approx 42^\circ$, and for a gold nanosphere with a diameter of 40 nm located 5 nm above a glass-water interface.  We use focus plane positions $z_{\rm foc}$ of (a) $-0.5$ $\mu$m, (b) 0, and (c) $+0.5$ $\mu$m.  The light blue bars inside the panels report the Michelson contrast, which is bound to values between zero and one.  The scale bar is $1\,\mu$m, the \iscat maps for $\theta=40^\circ$ have been multiplied by a factor of four for better visibility, and the maximal intensity $I_{\rm max}$ is the same in all panels.}\label{fig:brewster}
\end{figure*}

\begin{figure*}[t]
\centerline{\includegraphics[width=0.8\textwidth]{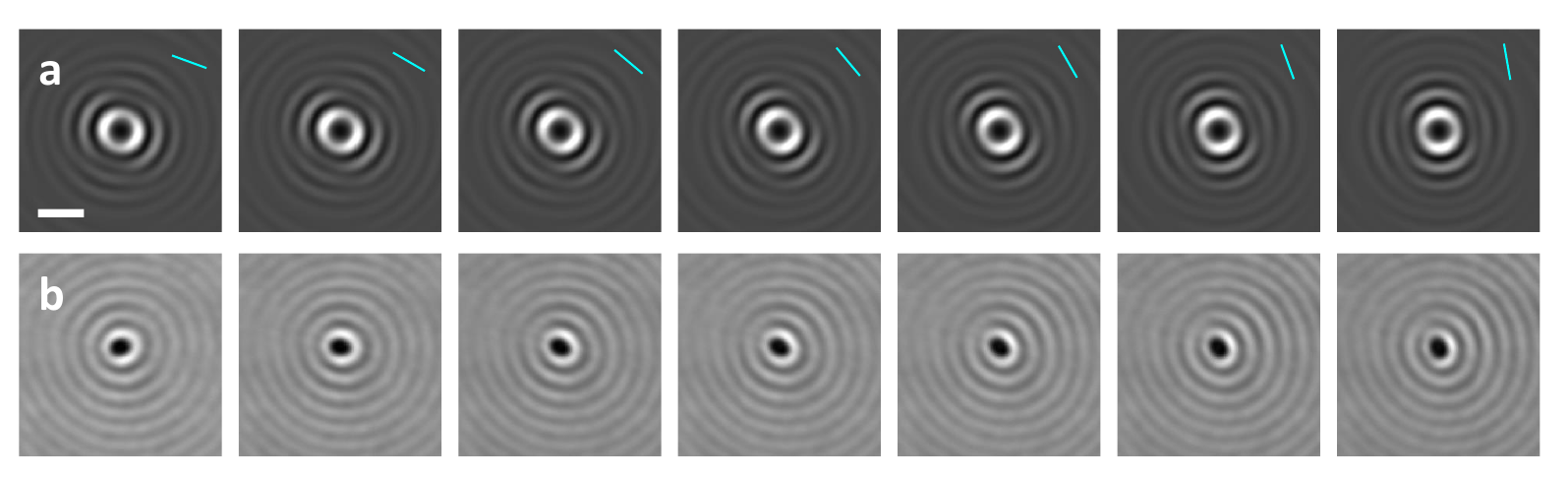}}
\caption{(a) Simulated and (b) experimental \iscat maps for a silver nanocube located on top of a glass-air interface (focus plane at $z=-0.4\,\mu$m).  The incoming light propagates in the positive $z$-direction, and the light polarization angles $\varphi_{\rm pol}$ are indicated in the upper right corners of the panels in (a).}\label{fig:cube}
\end{figure*}

Figure~\ref{fig:experiment} shows (a,b) simulated \iscat maps for a gold nanosphere with a diameter of 55 nm situated in air and 5 nm above a glass substrate.  In the simulations, we compute the background-subtracted interference between the secondary (reflected) incoming fields and the scattered fields via
\begin{eqnarray}\label{eq:iPSF}
  \mathcal{I}_{\rm sim}^\theta(x,y)&=& \left|\bm E_{\rm ref}'+\bm E_{\rm sca}'\right|^2
  -\left|\bm E_{\rm ref}'\right|^2 \nonumber\\ &=&
  \left|\bm E_{\rm sca}'\right|^2+
  2\left|\bm E_{\rm ref}'\right|\left|\bm E_{\rm sca}'\right|\cos\phi\,.
\end{eqnarray}
The angle of the incoming field with respect to the $+z$-axis is denoted with $\theta$, $\phi(x,y)$ is the phase between the reference and scattered fields. 
For sufficiently small particles, $\mathcal{I}_{\rm sim}$ is the interferometric point spread function i\textsc{psf} of~\cite{mahmoodabadi:20}. 

The angle $\theta$ of the incoming plane wave is (a) $14^\circ$ and (b) $-2^\circ$, respectively.  The different panels report the images at different focus plane positions (indicated on top) with respect to the interface position of the glass substrate at $z=0$.  We additionally obtained \iscat images $\mathcal{I}_{\rm exp}(x,y;z_{\rm exp})$ for various focus plane positions $z_{\rm exp}$.  For details of the experiment, see Appendix~\ref{sec:experiment}.  Note that in the experiment, the precise value of $z_{\rm exp}$ is unknown, and a fit to theoretical results is generally needed to obtain a reliable estimate for $z_{\rm exp}$.  In our work we simulated images $\mathcal{I}_{\rm sim}(x,y;z)$ for various focus plane positions $z_{\rm sim}$ and determined the experimental focus plane position $z_{\rm exp}$ from the highest correlation with the simulation results,
\begin{equation}
  \sup_{z\in z_{\rm sim}}  \Big[
  \mbox{corr}\Big(\mathcal{I}_{\rm sim}^\theta(x,y;z),\mathcal{I}_{\rm exp}^\theta(x,y;z_{\rm exp})\Big)\Big]
  \,,
\end{equation}
The correlation is computed through the \textsc{matlab} function \texttt{corr2}, the corresponding \textsc{matlab} files are provided in the Supporting Information.  Fig.~\ref{fig:experiment}(c,d) show the correlation maps between the simulated and experimental \iscat images. The red symbols indicate the highest correlation for the experimental data presented in (a*,b*). The images are in good agreement with the respective simulation results.

Interestingly, for off-axis illumination, as in a*, the correlation plot yields an unambiguous $z_{\rm exp}$ for each data set. This is not the case for the data in b*, which was recorded with on-axis illumination. Here the correlation plot shows two lines with a crossing in the focus plane. For a single \iscat image, it will thus be difficult to judge whether it is over-focused or under-focused.

Our simulations allow us to take a closer look at this behavior.  When plotting the simulations similar to those in Fig.~\ref{fig:experiment}(a) in the $xz$-plane for $y=0$ (see red dashed line), we obtain the \iscat maps shown in Fig.~\ref{fig:iPSF}.  A localized diffraction-limited spot when the focus plane coincides with the particle plane and characteristic interference features between the incoming and scattered waves are observed. In the colored panels of the figure, we plot the interference angle $\phi$, see Eq.~\eqref{eq:iPSF}, as extracted from our simulations.  The pronounced features in the \iscat maps can be traced back to the phase variations in the interference term of Eq.~\eqref{eq:iPSF}.  We again see that illumination at an angle leads to a clear directionality in the observed interference features, which explains the unique mapping observed in the correlation plots in Fig.~\ref{fig:experiment}. Panel (c) shows that a similar behavior can be found in \cobri.  Note that in the \cobri simulation for the excitation from above, we simply have to change the propagation direction of the incoming wave to $180^\circ$, resulting in a propagation into the negative $z$-direction.  The rest of the simulation remains identical.

Another advantage of off-axis illumination is illustrated in Fig.~\ref{fig:brewster}, where we simulate \iscat images with an illumination angle around the Brewster angle. We simulate a gold nanosphere (diameter 55 nm) at a glass-water interface.  For $x$-polarized light, no light will be reflected at the Brewster angle $\theta_B$.  This leads to a dark-field image where $\bm E_{\rm refl}^\downarrow=0$ and $\mathcal{I}=|\bm E_{\rm sca}'|^2$. For small deviations from $\theta_B$, the magnitude of the reference light, and thus the contrast in the \iscat image, can be tuned continuously.  This presents an alternative to other contrast-tuning modalities commonly used in \iscat, which are either based on polarization or on attenuators in the back-focal plane~\cite{Liebel2017, Cole2017, Cheng2019}.

In Fig.~\ref{fig:cube}, we present simulations and experiments for non-spherical particles, specifically for a silver nanocube with a side length of 100 nm.  In the simulations we use the dielectric function of~\cite{johnson:72}, details of the experiments can be found in Appendix~\ref{sec:experiment}.  The data were obtained with linear polarization at an angle $\varphi_{\rm pol}$ in the $xy$-plane.  Clearly, a rotation of $\varphi_{\rm pol}$ also rotates $\mathcal{I}$. The simulations, for which only the particle shape in the \textsc{{bem}} simulation had to be changed, and experiment show good agreement.  These results could not be easily attained with a simple dipole model, demonstrating the versatility of our numerical method. 

\subsection{Additional Features and Limitations}
Various \iscat modalities can be modeled with our toolbox. A prominent example is \iscat in a confocal scanning geometry, where the excitation beam is focused and scanned across the sample. Implementing this modality using our toolbox is described in the Supporting Information, where we also provide links to open-source demo files. 

Another example is \iscat with excitation light of limited longitudinal coherence, which prevents unwanted interference from reflections off interfaces other than the one close to the sample. This affects the interferometric point spread function.  It can be simulated by calculating the i\textsc{psf} for a number of wavelengths out of the spectrum of the light source. A weighted, incoherent sum of the simulated intensities will provide the final simulation result.   

We briefly comment on the accuracy and the limitations of our \bem approach.  The Stratton-Chu approach and the Galerkin implementation are known to give accurate results for sufficiently fine particle discretizations~\cite{buffa:03}.  The error due to the boundary discretization scales with $h^{\nicefrac 32}$, where $h$ is the mesh size.  As a rule of thumb, the ratio between $h$ and the effective wavelength in the medium should be at least of the order of $1:5$.  The \nanobem toolbox in its present form works best for nanoparticle discretizations with a few thousand boundary elements, say up to $10\,000$ elements, whereas runtimes for structures with more elements become prohibitively long.  The toolbox can currently only handle particles composed of materials with homogeneous material properties, which are separated by abrupt interfaces, but not particles with genuine inhomogeneous material properties.  See also the Supporting Information for a discussion of what has to be considered by users in setting up their own simulations.

\begin{figure*}[t]
\includegraphics[width=1.65\columnwidth]{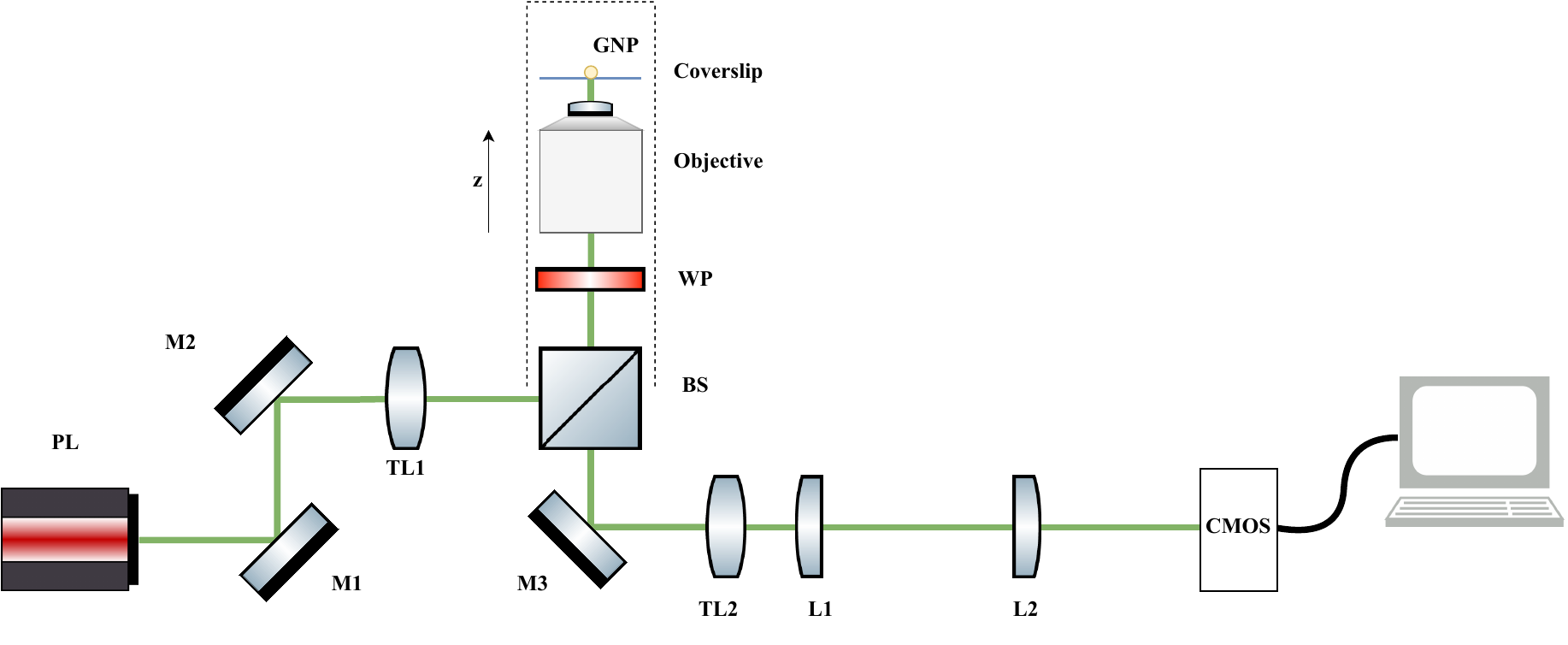}
\caption{Optomechanical components used in our experiments.  Silver-coated Thorlabs mirrors (M1, M2, M3), tube lenses TL1 with focal length $f_1 = 165$ mm, TL2, $f_2 = 165$ mm, lenses L1, L2 form an 4$f$-system; WP and BS are abbreviations for waveplate and beamsplitter respectively. For setup I: BS is polarizing beamsplitter, lenses L1, f3 = 150 mm , L2, f4 = 500 mm. For setup II: BS is non-polarizing beamsplitter and WP is not present, lenses L1, f3 = 75 mm , L2, f4 = 200 mm. Image composed with~\cite{franzen2006componentlibrary}. }
\label{fig:setup1}
\end{figure*}

\begin{table*}
\caption{Experimental parameters used for Figs. 6 and 9.  The propagation angles in setup I were determined by best correspondence with the simulations. The polarization is indicated for the light impinging on the particle.}
\label{tab:param}
\begin{tabularx}{2\columnwidth}{XlXX}
\hline\hline
Description & Symbol & Setup I & Setup II \\
\hline
Particle position & $z_{p}$ & on coverslip & on coverslip \\ 
Particle material and shape &  & gold sphere  & silver cube \\
Particle size & & 55 nm diameter & 100 nm cube length \\
Propagation angles & $\theta$ & $-2^{\circ}$, $14^{\circ}$ & $0^{\circ}$ \\
Wavelength & $\lambda$ & 517 nm & 450 nm \\
Refractive index object side  & $n$ & 1.5 & 1.5 \\
Refractive index image side & $ n' $ & 1.0 & 1.0 \\
Numerical aperture & NA & 1.3 &  1.42 \\
Magnification total system & $M_{\rm tot}$ & 133.33 & 160 \\
Pixel size & $pix_{\rm size}$ & 3.45 $\mu$m & 1.85 $\mu$m \\
Field of view & FOV & $3.9\times3.9\, \mu$m$^2$& $4.4\times4.4\, \mu$m$^2$ \\
Polarization & & circular & linear \\
\hline
\hline
\end{tabularx}
\end{table*}

\section{Summary}\label{sec:summary}

We have developed a simulation platform for interference microscopy, such as \iscat or \cobri, using a generic Maxwell solver based on the boundary element method.  The software, together with a detailed description, is provided in the Supporting Information.  Here presented the main building blocks of our simulation approach and provided several  examples. Experiments with gold nanospheres and silver nanocubes show good agreement with our simulation. The latter cannot be simulated with a simple dipole model. We found that \iscat contrast can be increased using an excitation close to the Brewster angle, demonstrating the use of our toolbox for exploratory purposes. Altogether, we hope that our software will be a valuable tool for researchers working in this field.

\section*{Acknowledgements}
This project has received funding from the European Union's Horizon 2020 research and innovation programme under grant agreement No 101017902 and No 758752.  The authors acknowledge the financial support by the University of Graz.

\begin{appendix}

\section{Experimental setup and parameters}\label{sec:experiment}

Our experimental data were recorded from two different \iscat setups and in the following we provide parameters and technical details of these.

\textbf{Setup I} is used for the data shown in Fig.~\ref{fig:experiment}.  Our light source is an ultrafast pulsed laser (Coherent Monaco) with a central wavelength $\lambda = 1035$ nm.  By frequency doubling before entering the \iscat setup, the resulting wavelength is $\lambda = 517.5$ nm.   We use an immersion oil objective (ZEISS, EC ``Plan-Neofluar'' 40x/1.30 Oil DiC M27) with 1.3 NA. The particles are diluted with distilled water and then applied to a glass coverslip. We let the mixture of GNP and water dry for 24 hours before using it. The coverslip with the nanoparticles (DNA-functionalized GNPs: Custom Oligo Conjugated Spherical Gold Nanoparticles for Zero Order Diagnostics from nanopartz, functionalized with a 21 nucleotide long DNA sequence) on top is mounted on a MadCityLabs nano translation stage, which enables us to scan the particle through the focus of the objective.  For the detection of the superposed fields, we use a CMOS camera (FLIR Blackfly BFSU3-50S5M).  We operate with the \textsc{matlab} image acquisition toolbox for image acquisition. Table~\ref{tab:param} summarizes further details.

\textbf{Setup II} is used for the data shown in Fig.~\ref{fig:cube}.  Our light source is an LED with a wavelength $\lambda = 450$ nm. The mode is cleaned by the polarisation maintaining fiber. The polarization state is then set by the fiber polarization controller (Thorlabs MPC320). It transforms the state of polarization of the laser output to any arbitrary state. A non-polarizing 10:90 cube beamsplitter is used in the setup. In this case 10$\%$ of the laser power is sent to the sample and 90$\%$ of the signal from the sample is transmitted to the CMOS camera (MER2-1220-32U3M, DAHENG IMAGING). No waveplate is needed for this setup. The state of polarization is measured by the polarimeter (Thorlabs PAX1000) mounted above the sample. The degree of polarization for all the measurements stayed above 90$\%$ and ellipticity was lower than  $0.4^{\circ}$. We use an immersion oil objective (Olympus PlanApo 60x with 1.42 NA). 

The particles are 100 nm silver nanocubes (NanoXact Silver Nanocubes--PVP) from nanocomposix. $20\, \mu$g/ml are applied to the hydrophilicly treated (plasma cleaned) coverslip for 10 minutes. The solution is then blown away. This procedure results in nanocubes being deposited on the glass surface. The data is acquired for a series of linear polarizations of the incoming field. For each angle of linear polarization $\phi_{pol}$, we record ten images with an exposure time of 7 ms. In post-processing, their mean is calculated. The resulting images are Fourier-filtered to remove the interference pattern caused by the side reflections in the setup. The background image corresponding to each state of polarisation is obtained by shifting the nanocube away from the field of view and taking the measurement. The background is then removed in post-processing. Table~\ref{tab:param} summarizes further details.

\section{Numerical implementation}\label{sec:simulation}

\begin{figure}[t]
\includegraphics[width=\columnwidth]{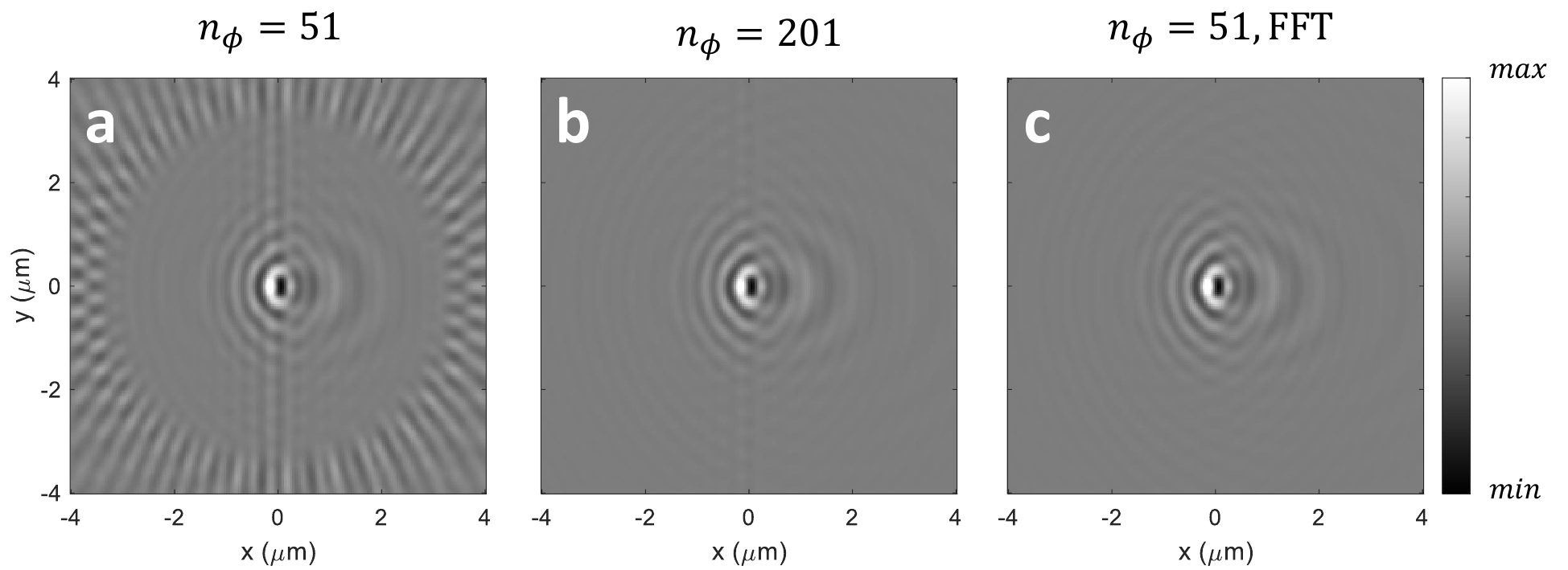}
\caption{Simulation results for interference between reflected incoming and scattered fields for a lens with ${\rm NA}=1.3$.  We use the Richards-Wolf approach and different angle discretizations, this is $n_\theta=50$ and (a) $n_\phi=51$, (b) $n_\phi=201$, and (c) $n_\phi=51$ together with the integration procedure of Eq.~\eqref{eq:bessel}.}
\label{fig:richardswolf}
\end{figure}

In this Appendix we provide some details about our implementation of the Richards-Wolf imaging procedure.  We first note that the most time-consuming part in our simulation approach is the solution of the \bem working equations.  We have thus currently refrained from implementing a full fast Fourier transform (\textsc{fft}) in the evaluation of the Richards-Wolf integration of Eq.~\eqref{eq:richards-wolf}, see~\cite{khadir:19,hohenester:20} for a more detailed discussion.  In principle, such an approach would not be overly complicated, and could be added to our software in the future, however, a more direct evaluation of the imaging integral in Eq.~\eqref{eq:richards-wolf} has the advantage that the far-field $(\phi,\theta)$ and image $(\rho,\varphi)$ coordinates can be chosen independently from each other, in contrast to \textsc{fft} where the two coordinate systems are directly linked.  Fig.~\ref{fig:richardswolf}(a) shows the simulated \iscat image for the setup depicted in Fig.~\ref{fig:bemsimulation} and for an equidistant discretization for the azimuthal and polar angles.  One observes the typical \iscat interference features in the center, but also numerical artifacts in the outer regions of the image.  These artifacts can be traced back to the exponential term in Eq.~\eqref{eq:richards-wolf}, which oscillates wildly for larger values of $\rho$.  Indeed, when increasing the number of azimuthal discretization points in panel (b) these artifacts start to disappear.  A computationally more efficient approach is to perform a \textsc{fft} transformation for the (equally discretized) azimuthal coordinate in the far-fields,
\begin{equation}
  \bm F(\theta,\phi)=\sum_m e^{im\phi}\tilde{\bm F}_m(\theta)\,.
\end{equation}
\noindent Each term of the Fourier series can then be integrated analytically~\cite[Eq.~(3.21)]{hohenester:20}
\begin{equation}\label{eq:bessel}
  \int_0^{2\pi}e^{im\phi}e^{i\lambda\cos(\phi-\varphi)}\,d\phi=2\pi i^m J_m(\lambda)e^{im\varphi}\,,
\end{equation}
where $J_m$ is the Bessel function of order $m$.  In particular for large values of $\lambda$, Eq.~\eqref{eq:bessel} gives accurate results for a much smaller number of discretization points, see panel (c), in comparison to the direct evaluation.  The evaluation can be additionally accelerated through truncation of the Fourier series, although this is not needed in most cases of interest.

\end{appendix}


\begin{widetext}

\setcounter{section}{4}
\section{Supporting Information}

\subsection{Getting started}

In this section, we briefly describe how to set up the toolbox and how to run simple \textsc{bem} simulations for interference microscopy.  To install the toolbox, one must add the path of the main directory \texttt{nanobemdir} of the \nanobem toolbox as well as the paths of all subdirectories to the \textsc{matlab} search path.  This can be done, for instance, through
\begin{code}
>> addpath(genpath(nanobemdir))
\end{code}
To set up the help pages, one must once change to the main directory of the \nanobem toolbox and run \texttt{makehelp}
\begin{code}
>> cd nanobemdir
>> makehelp
\end{code}
Once this is done, the help pages, which provide detailed information about the toolbox, are available in the \textsc{matlab} help browser under \texttt{Documentation>Supplemental Software}.  See also~\cite{hohenester.cpc:22,nanobem23} for further details.  In the \nanobem toolbox provided in the Supporting Information we have included the \verb!+optics! folder into the main directory of the toolbox.  The command \verb!addpath(genpath(nanobemdir))! then automatically makes all toolbox and \verb!+optics! functions available.  The demo programs accompanying this paper as well as a short documentation of the novel classes and functions is available in the help pages of the toolbox, see also Fig.~\ref{fig:toolbox}.  For the demo programs one simply has to run the \textsc{matlab} files.  

\begin{figure*}[t]
\centerline{\includegraphics[width=0.65\textwidth]{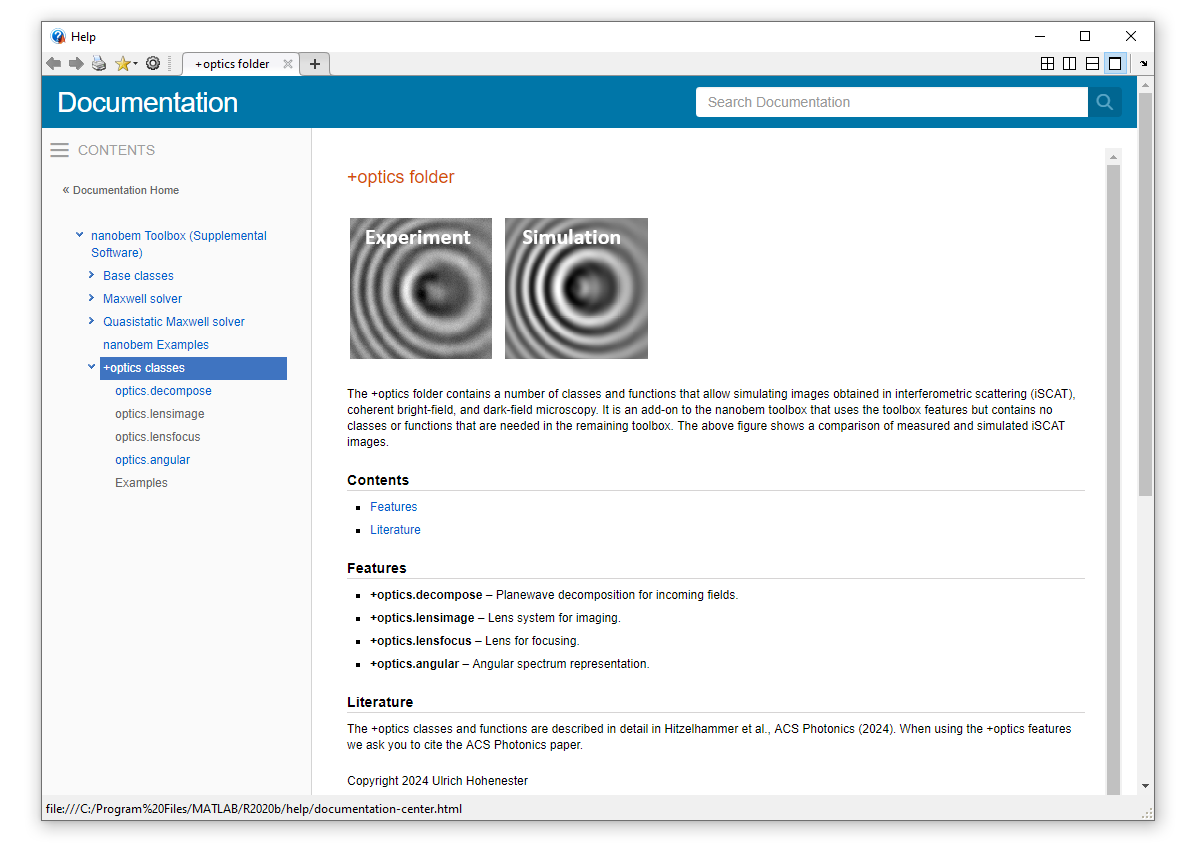}}
\caption{Help pages of the \nanobem toolbox which includes the \texttt{+optics} features for the simulation of coherence microscopies.}
\label{fig:toolbox}
\end{figure*}

\subsection{A simple example}

As a first example we consider a gold nanosphere above a glass substrate.  The example closely follows the demo file \texttt{demoiscat01.m}.  Each \bem simulation starts by defining the materials, the substrate or stratified medium, and the discretized particle boundary.

\begin{boxcode}
mat1 = Material( 1.50 ^ 2, 1 );                %  optical properties for glass
mat2 = Material( 1.33 ^ 2, 1 );                %  optical properties for water
mat3 = Material( epstable( 'gold.dat' ), 1 );  %  optical properties for gold
mat = [ mat1, mat2, mat3 ];                    %  unique material vector

layer = stratified.layerstructure( mat( 1 : 2 ), 0 );    %  layer structure
diameter = 50;                                           %  sphere diameter
p = trisphere( 144, diameter );                          %  discretized sphere boundary
p.verts( :, 3 ) = p.verts( :, 3 ) + 0.5 * diameter + 5;  %  shift nanosphere
tau = BoundaryEdge( mat, p, [ 3, 2 ] );                  %  boundary elements 
\end{boxcode}
In line 4 we define a unique material vector, and in line 6 we define a layer structure with the interface at $z=0$ and the material \verb!mat(1)! (glass) below the interface and \verb!mat(2)! (water) above.  In line 8 we generate a discretized sphere boundary with a diameter of 50 nm, which is shifted in line 9 to a position where the lowest point of the boundary is located 5 nm above the interface at $z=0$.  Finally, in line 10 we define the boundary elements of the discretized boundary, which can be submitted to the \bem solvers provided by the \nanobem toolbox.  We are now ready to define the incoming fields and to solve the \bem equations.

\begin{boxcode}[firstnumber=11]
k0 = 2 * pi / 520;        %  wavenumber of light in vacuum
pol = [ 1, 0, 0 ];        %  polarization of incoming light
dir = [ 0, 0, 1 ];        %  propagation direction
einc = optics.decompose( k0, pol, dir );   %  planewave decomposition
qinc = einc( tau, 'layer', layer );        %  inhomogeneities for incoming fields
bem = stratified.bemsolver( tau, layer, 'order', [] );  %  initialize BEM solver
sol = bem \ qinc;                                       %  solve BEM equations
\end{boxcode}
In lines 11--13 we define the wavenumber of light in vacuum, and the polarization and propagation direction of the incoming light.  In line 14 we set up an \verb!optics.decompose! object for a planewave decomposition of the incoming fields.  We will comment further below on the differences between this object and the \texttt{planewave} objects provided by the toolbox.  In line 15 we evaluate the inhomogeneities \verb!qinc! of the incoming fields in presence of the layer structure.  In this evaluation the secondary reflected and transmitted fields are automatically added.  Finally, in line 16 we set up the \bem solver for the stratified medium and finally solve the working equations of the \bem approach, to obtain a solution object \verb!sol! that stores the tangential electromagnetic fields at the particle boundary.

The main part of the \bem simulation is finished at this point.  What remains to be done is to compute the far-fields and to submit them to imaging.

\begin{boxcode}[firstnumber=18]
NA = 1.3;                               %  numerical aperture
air = Material( 1, 1 );                 %  optical properties for air
rot = optics.roty( 180 );               %  rotation matrix for imaging lens
lens = optics.lensimage( mat1, air, k0, NA, 'rot', rot );
far = farfields( sol, lens.dir );       %  farfields from BEM solution
refl = secondary( einc, layer, 'dir', 'down' );
\end{boxcode}
In line 18 we define the numerical aperture of the imaging lens, in line 19 we define the material properties on the image side of the objective (air), and in line 20 we initialize a rotation matrix ($180^\circ$ around $y$-axis) that rotates the optical axis from the $+z$ to the $-z$-direction.  In line 21 we set up an \verb!optics.lensimage! object that allows us to perform the Richards-Wolf imaging procedure.  When providing the rotation matrix to this object, all rotations of fields and propagation directions will be done automatically by the \verb!lens! object.  In line 22 the \bem solution and the \verb!lens! object work together: we compute the optical far-fields $\bm F(\hat{\bm r})$ of the \bem solution that propagate in the directions $\hat{\bm r}$ requested by the lens object.  Finally, in line 23 we compute the secondary fields of the incoming excitation that propagate downwards.  For the planewave excitation from below, the secondary fields are the fields reflected at the interface.  We are now ready to compute the imaged fields.

\begin{boxcode}[firstnumber=24]
x = 2000 * linspace( -1, 1, 201 );  %  image positions
z = 500;                            %  focus plane
focus = [ 0, 0, z ];                %  focus position
isca = efield( lens, far,  x, x, 'focus', focus );  %  scattered image fields 
iref = efield( lens, refl, x, x, 'focus', focus );  % incoming  image fields
isca = flip( isca, 1 );                   %  flip x-axis because of rotation matrix
iref = flip( iref, 1 );                   %  flip x-axis because of rotation matrix
im = dot( isca + iref, isca + iref, 3 );  %  interference image
\end{boxcode}
In line 24 we define the image field positions and in line 25 the focus plane.  In line 26 we define the focus position of the imaging system, and finally compute in lines 27, 28 the imaged fields.  In this computation we essentially solve the integrals entering the Richards-Wolf approach. Finally, we have to flip the images in the $x$-direction because of the rotation matrix in the \verb!lens! object.  In line 30 we compute the \iscat intensity $|\bm E_{\rm sca}+\bm E_{\rm ref}|^2$, which can be plotted and directly compared with experiment.  Interested users might like to consult and run the demo programs distributed with our software.

\subsection{Optics classes and functions}

The additional features needed for the simulation of interference microscopy are stored in the \verb!+optics! folder.

\begin{code}
+optics.decompose       %  planewave decomposition of fields
+optics.lensimage       %  imaging of fields through lenses
+optics.lensfocus       %  focusing of incoming fields through lens
+optics.angular         %  angular spectrum representation
\end{code}
None of the functions is optimized for speed, although they are usually sufficiently fast and designed to be versatile for the simulation of a wide range of different setups.  In the following we describe the various classes and functions in more detail.

\smallskip\ \\ \noindent\textbf{optics.decompose}\ \medskip

This class stores a planewave decomposition.

\begin{boxcode}
classdef decompose  
  %  Planewave decomposition.  
  properties
    k0      %  wavenumber of light in vacuum
    efield  %  electric field components
    dir     %  propagation directions
  end
end
\end{boxcode}
\verb!k0! is the wavenumber of light in vacuum, and \verb!efield! and \verb!dir! are the field amplitudes $\bm\epsilon$ and wavevector direction $\hat{\bm k}$, respectively.  In case of a single plane wave \verb!efield! and \verb!dir! are single vectors.
\begin{code}
%  plane wave with x-polarization and z-propagation
field = optics.decompose( k0, [ 1, 0, 0 ], [ 0, 0, 1 ] );
\end{code}
When \verb!efield!, \verb!dir! are \verb!n!$\times$\verb!3! arrays, the electric fields are computed from
\begin{equation*}
  \bm E(r)=\sum_i \exp({i\bm k_i\cdot\bm r})\bm\epsilon_i\,.
\end{equation*}
Note that this summation over $i$ differs from the \verb!planewave! objects of the \nanobem toolbox, which allows us to obtain the response of a nanoparticle for various planewave excitations simultaneously.  In contrast, with the \verb!optics.decompose! objects we can only run simulations for a single (albeit possibly more complicated) incoming field.  This functionality has certain advantages in the context of interference microscopy, in particular when it comes to the focused excitations described below.  Once the object has been initialized, it can be used in different ways.  The secondary fields, these are the reflected or transmitted fields of the stratified medium, can be computed from

\begin{code}
field2 = secondary( fields, layer, 'dir', dir );  
\end{code}
Here \verb!layer! is the layer structure of the stratified medium, and the propagation direction must be chosen as either \verb!'up'! or \verb!'down'!.  To evaluate the fields at given positions \verb!pos!, we first have to place the position points in the photonic environment before submitting them to the \verb!fields! function.  For a homogeneous medium this can be done with
\begin{code}
pts = Point( tau, pos );          %  place position(s) in photonic environment
[ e, h ] = fields( field, pts );  %  evaluate planewave decomposition
\end{code}
For a stratified medium the code has to be changed to
\begin{code}
pts = stratified.Point( layer, tau, pos );
[ e, h ] = fields( field, pts, 'layer', layer );
\end{code}
Note that in this evaluation all secondary fields originating from reflections or transmissions at the interfaces of the stratified medium are added automatically.  The inhomogeneity for the \bem solver can be computed from
\begin{code}
qinc = eval( field, tau );                   %  homogeneous embedding medium
qinc = eval( field, tau, 'layer', layer );   %  particle in stratified medium
\end{code}

\smallskip\ \\ \noindent\textbf{optics.lensimage}\ \medskip

This class allows to simulate imaging of far-fields and planewave decompositions using the Richards-Wolf approach.  

\begin{boxcode}
classdef lensimage  
  %  Imaging lens, Hohenester "Nano and Quantum Optics", Eq. (3.10).  
  properties
    mat1       %  material properties on object side
    mat2       %  material properties on image  side
    k0         %  wavenumber of light in vacuum
    NA         %  numerical aperture
    dir        %  directions for far-fields
    mcut       %  cutoff for angular degree
    rot        %  rotation matrix for optical axis
    backfocal  %  field manipulation in backfocal plane
  end
end
\end{boxcode}
The class can be initialzed through
\begin{code}
lens = optics.lensimage( mat1, mat2, k0, NA,  ...
          'rot', rot, 'backfocal', fun, 'nphi' nphi, 'ntheta', ntheta, 'mcut', mcut );
\end{code}
The properties in the second line are optional.  If they are not specified by the user, the default values are used.  The meaning of \verb!mat1!, \verb!mat2!, \verb!k0! and \verb!NA! is self-explanatory.  \verb!rot! is an additional rotation matrix that allows rotating the optical axis from the $+z$-direction into other directions, as described in more detail below.  \verb!backfocal! is a user-defined function that can be used to manipulate the electromagnetic fields in the backfocal plane, for instance to mimic the effect of a quarter-wave plate.  \verb!nphi=51! and \verb!ntheta=50! are the discretiaztions of the Gaussian reference sphere used in the Richards-Wolf approach, and \verb!mcut! allows for an optional truncation of the partial Fourier transform of the optical far-fields, as discussed in the main text.  Upon initialization, \verb!lens.dir! defines the (rotated) directions in which the optical far-fields have to be computed.  With a solution \verb!sol! of the \bem solver provided by the toolbox, these far-fields can be computed and plotted through
\begin{code}
far = farfields( sol, lens.dir );  %  compute farfields in directions lens.dir
plot( lens, far );                 %  plot farfields on Gaussian reference sphere
\end{code}
Once the farfields \verb!far! or planewave decompositions \verb!ref! of the  secondary incoming fields are at hand, they can be submitted to the imaging of the Richards-Wolf approach through
\begin{code}
isca = efield( lens, far, x, y, 'focus', focus );  %  imaging of farfields
iref = efield( lens, ref, x, y, 'focus', focus );  %  imaging of planewave decomposition
\end{code}
Here \verb!x!, \verb!y! are the image positions, and \verb!focus! is the position of the focus spot, as will be described in more detail further below.  \verb!isca! and \verb!iref! are arrays of size \verb!nx!$\times$\verb!ny!$\times$\verb!3!.

\smallskip\ \\ \noindent\textbf{optics.lensfocus}\ \medskip

This class allows to simulate focusing of an incoming laser field, see Eq.~(3.13) of~\cite{hohenester:20}.  The class works closely together with \verb!optics.decompose! objects for planewave decompositions.

\begin{boxcode}
classdef lensfocus
  %  Imaging lenses, Hohenester "Nano and Quantum Optics", Eq. (3.13).  
  properties
    mat     %  material properties on focus side
    k0      %  wavenumber of light in vacuum 
    NA      %  numerical aperture
    rot     %  rotation matrix for optical axis
  end
end
\end{boxcode}
The class can be initialzed through
\begin{code}
lens = optics.lensfocus( mat, k0, NA, 'rot', rot 'nphi' nphi, 'ntheta', ntheta );
\end{code}
\verb!mat! is the material on the focus side where the fields are computed, \verb!k0! is the wavenumber of light in vacuum, and \verb!NA! is the numerical aperture.  Users can also provide a rotation matrix \verb!rot! that rotates the focused fields.  The optional parameters \verb!nphi=31! and \verb!ntheta=30! control the discretization of the Gaussian reference sphere that is used to simulate focusing.  After initialization, the user must provide the incoming fields impinging on the Gaussian reference sphere (with unit radius, the cutoff radius for a given NA is \verb!lens.rad!) at the positions
\begin{code}
[ x, y ] = deal( lens.x, lens.y );           %  Cartesian coordinates
[ phi, rho ] = deal( lens.phi, lens.rho );   %  polar coordinates
\end{code}
For instance, for an incoming Gauss-Laguerre beam with a topological charge of $m=1$ and $x$-polarization we use
\begin{code}
e = normpdf( lens.rho, 0, 0.5 ) .* exp( 1i * lens.phi );
e = e( : ) * [ 1, 0, 0 ];
\end{code}
Once the fields impinging on the Gaussian reference sphere are determined, the focused fields are obtained from
\begin{code}
field = eval( lens, e );
field = eval( lens, e, 'focus', focus );  %  user-defined focus position in medium
\end{code}
\verb!field! is an \verb!optics.decompose! object that can be evaluated (for unbounded and stratified media) as previously described.  We have refrained from analytic integrations for the azimuthal angle, see Eq.~(3.21) of~\cite{hohenester:20}.  This makes the programs probably slower and less accurate, users might like to play with \verb!nphi!, \verb!ntheta! that control the degree of discretization of the reference sphere.  On the other hand, the present implementation is relatively flexible and allows for the simulation of a wealth of focused excitations.  

If needed, the fields of a planewave decomposition can be rotated by defining a rotation matrix \verb!rot! and applying it to the \verb!optics.decompose! object through
\begin{code}
field = rot * field;     %  apply rotation matrix from lhs
field = field * rot .';  %  apply rotation matrix from rhs
\end{code}
To shift the fields by a vector \verb!pos!, we have to compute the shift phaese and multiply them to the planewave decomposition (\verb!k! is the wavenumber in the medium)
\begin{code}
shift = exp( - 1i * k * field.dir * pos .' );   %  shift phases
field = field .* shift;                         %  multiply fields with shift phases
\end{code}
Alternatively, we can use the \verb!transform! function
\begin{code}
obj = transform( obj, 'rot', rot );
obj = transform( obj, 'shift', pos, 'mat', mat );
\end{code}
to rotate the fields or shift them by the vector \verb!shift! in a medium with the material properties \verb!mat!.  Note that in most cases of interest the rotation and shift operations are performed automatically in the \verb!optics.lensimage! and \verb!object.lensfocus! objects.

\smallskip\ \\ \noindent\textbf{optics.angular}\ \medskip

This function allows us to perform an angular spectrum decomposition, see Eq.~(3.8) of~\cite{hohenester:20}, and is mainly implemented for testing purposes.  Suppose that we know the fields \verb!e(x,y,1:3)! in a plane \verb!z! at the positions of a rectangular grid \verb!(x,y)!.  We assume that the fields are localized inside the grid \verb!(x,y)! and are approximately zero outside the grid.  We can then compute the far-fields using the angular spectrum representation,
\begin{code}
far = angular( mat, k0, dir, e, x, y, z, 'nmax', nmax );
\end{code}
\verb!mat! holds the material parameters of the embedding medium, \verb!k0! is the wavenumber in free space, and \verb!dir! are the propagation directions of the far-fields $\bm F(\hat{\bm r})$.  \verb!z=0! is the plane where the fields have been computed.  The function might require a lot of memory and the computation is broken up into a loop where in each iteration \verb!nmax=1000! directions are computed at a time.  This value should suffice for most cases of interest, but can be reduced if memory becomes an issue.  The output of the function are the far-fields \verb!far! with propagation directions \verb!dir!.

\subsection{Demo programs}

We provide a few demo programs that demonstrate the working principle of the software.

\begin{description}
\item[\texttt{\bf demooptics01.m}]  Focusing of incoming laser beam for user-defined focus position and rotated optical axis.

\item[\texttt{\bf demooptics02.m}]  Imaging of point dipoles for different focus planes and for imaging system with optical axis along $+z$, see Fig.~\ref{fig:lens01}(a--c).

\item[\texttt{\bf demooptics03.m}]   Imaging of point dipoles for different focus planes and for imaging system with optical axis along $-z$, see Fig.~\ref{fig:lens01}(a*--c*).

\item[\texttt{\bf demoiscat01.m}] Computes the \iscat images for a gold nanosphere (50 nm diameter, embedded in water) situated 5 nm above a glass substrate.  By changing the numbers in the \verb!switch! block at the beginning of the program one can simulate different setups, this is (2) coated, (3) dielectric, and (4) coupled nanospheres.

\item[\texttt{\bf demoiscat02.m}] Same as \verb!demoiscat01.m! for single gold nanosphere and varying focus planes.

\item[\texttt{\bf demoiscat03.m}]  Same as \verb!demoiscat02.m!, we additonally compare with the results of a simplified dipole model.

\item[\texttt{\bf demoiscat04.m}]  Same as \verb!demoiscat02.m! but with a quarter-wave plate in the excitation and detection path.

\item[\texttt{\bf demoiscat05.m}]  Compute \iscat images for different focal plane positions and display images together with an interactive slider.

\item[\texttt{\bf demoiscat06.m}]  Compute \iscat images for comparison with Mahmoodabadi et al., Opt. Express 28, 25969 (2020), Fig. 3(b).

\item[\texttt{\bf demofield01.m}]  Computes the electric field maps for the different setups described above.  

\item[\texttt{\bf democorr01.m}]  Example of how to obtain the optimal focus plane for an experimental image.  In this demo file we also show how to account for a quarter-wave plate.

\item[\texttt{\bf demofocus01.m}]  Compute the focus fields for an incoming Laguerre-Gauss beam.  We also image the focus fields using either the planewave decomposition or the far-fields obtained from the angular spectrum representation.  Users might like to compare the energy transported by the waves through the focus (\verb!P1!) and image (\verb!P2!) plane, which should be the same apart from numerical inaccuracies.
\end{description}

\begin{figure}[t]
\centerline{\includegraphics[width=0.45\columnwidth]{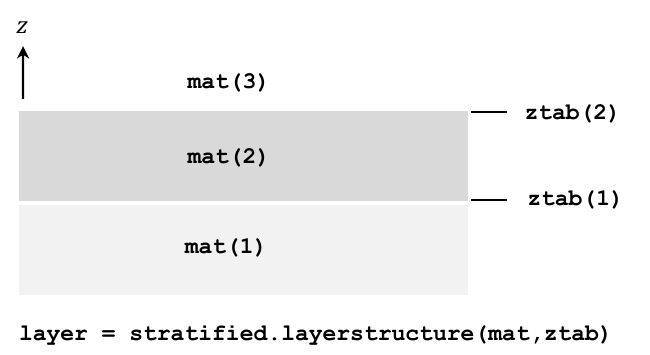}}
\caption{Definition of layer structure with material vector \texttt{mat} and interface positions \texttt{ztab}.}
\label{fig:layer}
\end{figure}

\subsection{Short list of How-tos}

\begin{figure*}[t]
\centerline{\includegraphics[width=0.6\columnwidth]{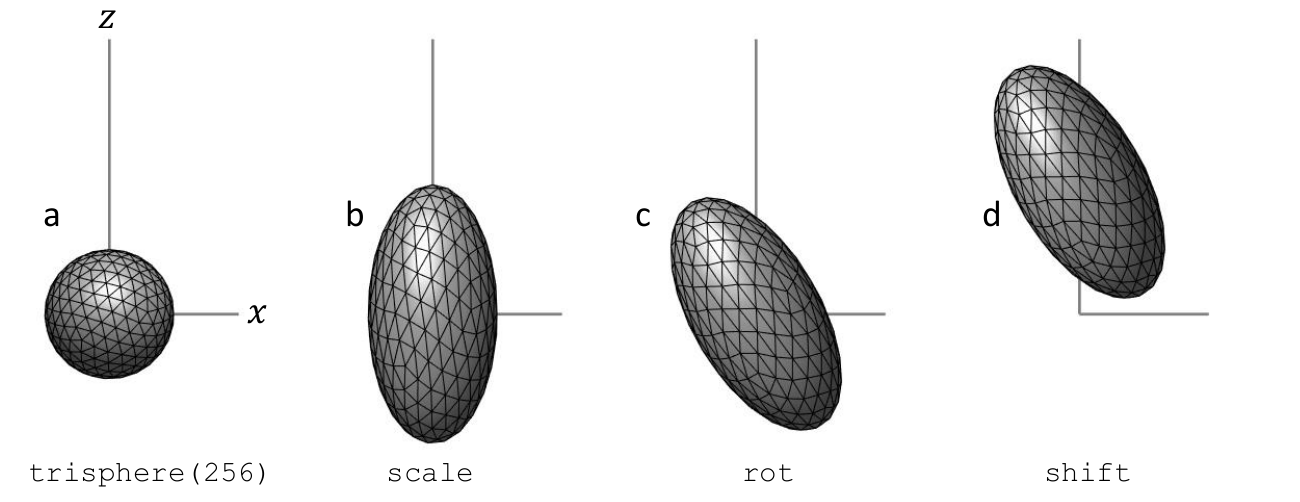}}
\caption{Nanoparticle boundary for (a) sphere and (b--d) transformed boundaries.}
\label{fig:particle}
\end{figure*}

In this section we discuss a few how-tos that might help users to set up their own \iscat simulations.

\medskip

\noindent\textbf{How to define rotation matrices?}  
\textsc{matlab} provides a few commands for setting up rotation matrices, as needed for the rotation of particles or the optical axis of lenses.  These include \verb!eul2rotm!, \verb!rotx!, \verb!roty!, and \verb!rotz!.  Additionally, we provide the three functions
\begin{code}
rot = optics.rotx( t );   %  rotation around x-axis by angle t (deg)
rot = optics.roty( t );   %  rotation around y-axis by angle t (deg)
rot = optics.rotz( t );   %  rotation around z-axis by angle t (deg)
\end{code}

\noindent\textbf{How to set up layer systems?}  
To set up a layer structure in the \nanobem toolbox, we have to define the material properties of the different media (in ascending order with respect to $z$) together with the interface positions (in ascending order with respect to $z$).  See also Fig.~\ref{fig:layer}.  We recommend using a unique material vector including all possible materials used in the simulations, with the first $n$ entries corresponding to the materials of the stratified medium with $n$ layers.  The remaining material entries are then used for the nanoparticles, as discussed above.  The stratified medium is then set up with
\begin{code}
layer = stratified.layerstructure( mat( 1 : n ), ztab( 1 : n - 1 ) );
\end{code}

\noindent\textbf{How to set up particles?}
The \nanobem toolbox provides a number of functions for defining the nanoparticle boundaries.  As an example, in Fig.~\ref{fig:particle}(a) we show a nanosphere with a given diameter that is produced with
\begin{code}
p = trisphere( 256, diameter );
\end{code}
The nanoparticle boundary can then be transformed using the functions
\begin{code}[numbers=left]
p = transform( p, 'scale', [ 1, 1, 2 ] );
p = transform( p, 'rot', optics.roty( 30 ) );
p = transform( p, 'shift', [ 0, 0, - min( p.verts( :, 3 ) ) + gap ] );
\end{code}
In line 1 we scale the particle along the three axes of the Cartesian coordinate system, in line 2 we rotate the particle by providing a rotation matrix, and in line 3 we shift the particle to a position where the lowest point of the boundary is located a distance \verb!gap! above the plane $z=0$.  With the latter command one can place nanoparticles above or below the interfaces of a stratified medium.  Rather than using the \verb!transform! commands one can also directly manipulate the vertices of the \verb!particle! object.
\begin{code}[numbers=left]
p.verts = p.verts .* [ 1, 1, 2 ];
p.verts = p.verts * optics.roty( 30 ) .';
p.verts = p.verts + [ 0, 0, - min( p.verts( :, 3 ) ) + gap ];
\end{code}
Note that in line 2 we use the transpose of the rotation matrix because we multiply it from the right-hand side on the vertices array \verb!p.verts(:,1:3)!.

\medskip

\begin{figure*}[t]
\includegraphics[width=0.8\columnwidth]{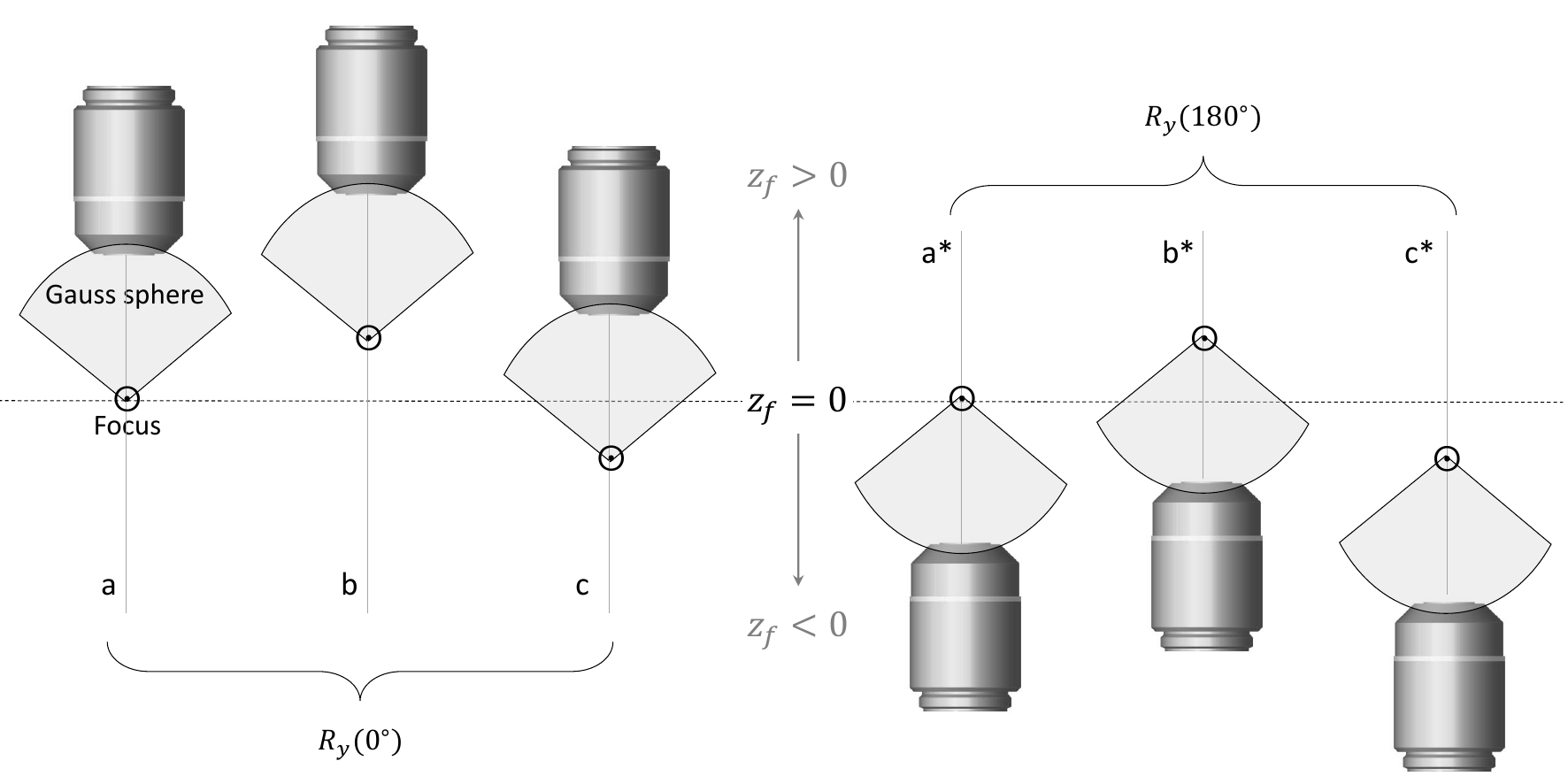}
\caption{Rotation of optical axis and user-defined focus position \texttt{[0,0,zf]} for imaging lenses.}
\label{fig:lens01}
\end{figure*}

\noindent\textbf{How to define the position and orientation of lens objects?}  On default the optical axis of the \verb!optics.lensimage! object points in the $+z$-direction and the focus point is located at \verb![0,0,0]!.  In order to rotate the optical axis we can specify in the initialization a rotation matrix \verb!rot!,
\begin{code}
lens = optics.lensimage( mat1, mat2, k0, NA, 'rot', rot );
\end{code}
To set a user-defined focus point (defined with respect to the center of the coordinate system of the \textsc{bem} simulations), we can provide the position of the focus point \verb!focus! in the evaluation of the image fields
\begin{code}
im = efield( lens, field, x, x, 'focus', focus );  %  user-defined focus position
\end{code}
This command works for both far-fields and planewave decompositions.  Through this command the focus point is moved to position \verb!focus! of the \textsc{bem} coordinate system, as shown in Fig.~\ref{fig:lens01} for \verb!focus=[0,0,zf]!.

A similar procedure can be used for the focusing lens.  In the initialization we can pass a rotation matrix \verb!rot! and in the evaluation the focus position \verb!focus!,
,
\begin{code}[numbers=left]
lens = optics.lensfocus( mat1, mat2, k0, NA, 'rot', rot );
field = eval( lens, efield, 'focus', focus );
\end{code}

\noindent\textbf{How to manipulate fields in the backfocal plane?}  Sometimes one might like to manipulate the fields in the backfocal plane of the imaging system, e.g. to mimic the effect of a quarter-wave plate.  We provide a function that gives the Jones matrix for a quarter-wave plate
\begin{code}
quarter = optics.quarterplate( t );  %  rotation angle in degrees
\end{code}
\verb!quarter! is a $3\times 3$ matrix for the optical axis along $z$ and \verb!t! is the rotation angle of the plate in degrees.  With \verb!optics.quarterplate(45)! we then get a quarter-wave plate that transforms an incoming field with linear polarization along $x$ into a field with circular polarization.  To modify the fields in the backfocal plane, we must provide a user-defined function \verb!fun! that manipulates the fields according to \verb!field=fun(field)!.  For the Jones matrix \verb!quarter! and the imaging lens object \verb!lens! we then set
\begin{code}
lens.backfocal = @( field ) field * quarter .';
\end{code}
Once the function is set, the field manipulations in the backfocal plane are performed automatically.  Note that we use the transpose of the Jones matrix because we apply the matrix from the right-hand side on the array \verb!field(:,1:3)!.  

To apply the same transformations for a focusing lens \verb!lens!, we can use the same function to manipulate the incoming fields before crossing the Gaussian reference sphere,
\begin{code}
efield = fun( efield );        %  apply transformation to fields before Gauss sphere
field = eval( lens, efield );  %  focused fields
\end{code}

\noindent\textbf{How to manipulate a single plane wave?}  To rotate, shift, and transform a plane wave we can proceed as follows:
\begin{code}[numbers=left]
pol = fun( pol );
field = optics.decompose( pol, dir );
field = transform( field, 'rot', rot );
field = transform( field, 'shift', shift, 'mat', mat );
\end{code} 
In line 1 we apply a transformation function to the field component of the plane wave, which is initialized in line 2.  In line 3 we rotate the fields and in line 4 we propagate the fields by the vector \verb!shift! in a medium with material properties \verb!mat!.

\subsection{Frequently asked questions}

\begin{description}
\item[\bf Which units does the toolbox use?] The \nanobem toolbox uses nanometers for length and the free-space wavenumber $k_0=\frac{2\pi}\lambda$ for frequencies.  The electric field is given in units of the incoming fields, the magnetic field is given in units of $Z_0\bm H$, where $Z_0$ is the free-space impedance, such that $\bm E$ and $Z_0\bm H$ have the same dimensions.

\item[\bf Which coordinate systems are used by the toolbox?] \

\begin{enumerate}
\item In the coordinate system of the \nanobem toolbox one specifies the nanoparticle boundaries with respect to a user-defined origin, which is usually chosen such that $z=0$ corresponds to the substrate interface.

\item In the coordinate system of the \texttt{optics.lensimage} object the optical axis is parallel to the $z$-axis and the the origin of the coordinate system is the focus point of the first Gaussian reference sphere.

\item In the coordinate system of the \texttt{optics.lensfocus} object the optical axis is parallel to the $z$-axis and the the origin of the coordinate system is the focus point of the Gaussian reference sphere.
\end{enumerate}
To align the coordinate systems we can rotate the optical axis and define the focus position \verb!focus! using the procedures discussed above.

\item[\bf How to compute optical forces?] The computation of optical forces through Maxwell's stress tensor is implemented in the toolbox.  For a \bem solution object \verb!sol! one obtains the optical force \verb!ftot! and torque \verb!ntot! through
\begin{code}
[ ftot, ntot, frc ] = optforce( sol );
[ ftot, ntot, frc ] = optforce( sol, 'imat', imat, 'ind', ind, );
\end{code}
\verb!imat=1! is the index to the material in which the nanoparticle is embedded, and \verb!ind! is an optional index for selected boundary elements.  \verb!frc! is the force component for each boundary element.

\item[\bf How to choose the best discretization?] This is a difficult issue.  Quite generally, within our \bem implementation it is guaranteed that the computational solution converges to the true solution for sufficiently fine meshes, and also particles with sharp edges or directly placed on interfaces of a stratified medium can be safely simulated.  It is usually a good idea to run simulations with an increasing number of boundary elements and to inspect the simulation results.  The final choice of fast runtimes (coarse discretizations) versus highly accurate results (fine discretizations) is left to the users, although in many cases even coarse discretizations will already give sufficiently accurate results.

\item[\bf How to speed up simulations for multilayer systems?]
For stratified media with more than one interface the simulations can be sometimes very slow.  This is usually because of an inefficient evaluation of the Sommerfeld integrals, see discussion in~\cite{nanobem23}.  It often helps to use in the initialization of the \bem solver the options
\begin{code}
op = odeset( 'AbsTol', 1e-5, 'InitialStep', 1e-3 );
bem = stratified.bemsolver( tau, layer, 'op', op, 'semi1', 10 );
\end{code}
Users are advised to play with the value of \verb!semi1! to find the best performance of the code.  \verb!AbsTol! controls the accuracy of the \textsc{matlab} \texttt{ode45} integration, however, the value should not be chosen too small because otherwise the simulations get inaccurate.

\item[\bf How to speed up simulations for many particle positions?]
To run simulations for various particle positions one has to introduce a loop over the different positions, set up the discretized particle boundary and the \bem solver, and compute the far-fields for each position separately.  This can result in long simulation times.  If the particle is sufficiently far away from the interfaces of the stratified medium and nearfield couplings play a minor role, there exists a more simple but approximate approach.  We solve the \bem equations for a homogeneous medium and account for the shifting of the particle by evaluating the incoming fields at the shifted particle positions.  The solution object is then converted to a \verb!stratified.solution! vector, which can be used in the same manner as previously described.
\begin{code}[numbers=left]
bem = galerkin.bemsolver( tau, 'order', [] );  %  BEM solver for homogeneous medium
einc = optics.decompose( k0, pol, dir );       %  primary excitation

for iz = 1 : numel( zout )
  qinc = einc( tau, 'layer', layer, 'shift', [ 0, 0, zout( iz ) ] );  
  [ sol, bem ] = solve( bem, qinc );           %  SOL w/o nearfield couplings
  sol = stratified.solution( qinc.tau, k0, sol.e, sol.h );
  sol.layer = layer;                           %  approximate SOL for layer
  ...
end
\end{code}
In line 5 we evaluate the incoming fields at the shifted particle position, where \verb!zout! is an array of the vertical shift positions.  Note that \verb!qinc.tau! is an array of the shifted boundary elements.  With the calling sequence in line 6 we ensure that the \bem solver is evaluated only once.  Quite generally, this approach is much faster and justified when nearfield couplings of the particle to the stratified medium are of minor importance.  Whether this is the case should be examined for a few shift positions through comparison with the results of the full simulations.

\item[\bf What can go wrong in simulation?]  We have done our best to make the \nanobem toolbox as stable as possible.  Nevertheless, a number of things might go wrong.  We advise users to carefully check whether the boundaries of the nanoparticles are defined properly, in particular with respect to the orientation of the boundary and the choice of inner and outer materials.  For stratified media one should be also careful about the reflected Green's functions, which can be plotted for testing as described in~\cite{nanobem23}.  In case of other unforeseen errors, contact one of the authors with a detailed description of the problem.

\end{description}
\end{widetext}

\end{document}